\documentclass[aos,preprint]{imsart}
\setattribute{journal}{name}{}

\usepackage{graphicx}
\usepackage{subfig}
\usepackage{color}
\usepackage{url}

\usepackage{amsmath}
\usepackage{amssymb}
\usepackage{bbm}
\usepackage{multirow}
\usepackage{fullpage}
\usepackage[round]{natbib}

\startlocaldefs
\usepackage{macros}
\endlocaldefs

\begin{document}

\begin{frontmatter}
\title{Bayesian Structured Sparsity from Gaussian Fields}
\runtitle{Bayesian Structured Sparsity from Gaussian Fields}

\begin{aug}
\author{\fnms{Barbara E.} \snm{Engelhardt}\ead[label=e1]{barbara.engelhardt@duke.edu}}
\and
\author{\fnms{Ryan P.} \snm{Adams}\ead[label=e2]{rpa@seas.harvard.edu}}
\affiliation{Duke University and Harvard University}

\runauthor{B. E. Engelhardt and R. P. Adams}

\address{Barbara E. Engelhardt\\
Department of Biostatistics \& Bioinformatics \\
Institute for Genomic Science \& Policy \\
Department of Statistical Science\\
Duke University\\
Durham, NC\\
\printead{e1}}

\address{Ryan P. Adams\\
School of Engineering and Applied Sciences\\
Harvard University\\
Cambridge, MA\\
\printead{e2}}

\end{aug}

\begin{abstract}
Substantial research on structured sparsity has contributed to
analysis of many different applications.  However, there have been few
Bayesian procedures among this work.  Here, we develop a Bayesian
model for structured sparsity that uses a Gaussian process (GP) to
share parameters of the sparsity-inducing prior in proportion to
feature similarity as defined by an arbitrary positive definite
kernel. For linear regression, this sparsity-inducing prior on
regression coefficients is a relaxation of the canonical
spike-and-slab prior that flattens the mixture model into a scale
mixture of normals. This prior retains the explicit posterior
probability on inclusion parameters---now with GP probit prior
distributions---but enables tractable computation via elliptical slice
sampling for the latent Gaussian field. We motivate development of
this prior using the genomic application of association mapping, or
identifying genetic variants associated with a continuous trait. Our
Bayesian structured sparsity model produced sparse results with
substantially improved sensitivity and precision relative to
comparable methods. Through simulations, we show that three properties
are key to this improvement: i) modeling structure in the covariates,
ii) significance testing using the posterior probabilities of
inclusion, and iii) model averaging. We present results from applying
this model to a large genomic dataset to demonstrate computational
tractability.
\end{abstract}
\end{frontmatter}

\section{Introduction and Motivation}
\label{sec:sparsity}
Sparsity is an important tool in applied statistics from three
perspectives.  First, in the settings where there are many more
features than samples (so-called~${p \gg n}$ problems), employing a
sparsity-inducing prior, or penalty term, has proven to be effective
for regularization.  For regression, the likelihood-maximizing
parameters in the unregularized~${p\gg n}$ setting correspond to a
continuum of solutions in a high-dimensional linear subspace.  Many of
these solutions will result in \emph{overfitting}, in which the
samples are well-captured by the parameters, but the model generalizes
poorly for out-of-sample data.  Overfitting is typically avoided by
penalizing parameters, and in the Bayesian setting this corresponds to
specifying a prior for the coefficients.  In high dimensional
problems, penalties that remove features by setting their
contributions to $0$ are ideal: rather than choosing the
likelihood-maximizing parameters with the smallest Euclidean norm,
overfitting may be overcome through \emph{model selection}, or
choosing feature set with the smallest number of elements.  Model
selection is performed for penalized regression by minimizing the
$\ell_0$ norm of the feature inclusion parameters.  In practice,
however, this is an exponentially large search space with $2^p$
possible solutions; convex relaxations provide popular and effective
approximate objectives and enable computational tractability.
Continuous Bayesian sparsity-inducing priors, such as the
double-exponential~\citep{Hans2009} and the
horseshoe~\citep{Carvalho2009}, as well as explicit penalization
approaches such as $\ell_1$ (Lasso)~\citep{Tibshirani1996} and elastic
net~\citep{zou2005regularization}, have been used effectively for such
relaxation.  Mapping these sparse feature parameters in the relaxed
space back to the corners of a $p$-hypercube, in order to determine
which features are included and which are excluded, must be performed
thoughtfully, however.

Second, sparse priors often enable computational tractability by
explicitly modeling a lower dimensional feature
space~\citep{Tropp2010}.  Such computational savings have often been
difficult to realize in practice; for example, the LARS algorithm for
$\ell_1$ regularized regression considers all features at each
iteration~\citep{Efron2004}. One explanation for this behavior is
that, in general, these methods were not optimized for the $p\gg n$
setting. Here we take advantage of sparsity to improve tractability in
parameter estimation.

Third, sparsity, and model selection more generally, are crucial to
the scientific goal of discovering which features are useful for our
statistical task and which may be safely
ignored~\citep{OHara2009,Breiman2001}.  In problems across the sciences
and beyond, estimating the relative contribution of each feature is
often less important than estimating whether or not a feature
contributes at all~\citep{Petretto2010}.  The downstream value of
selecting a small subset of features is in creating a small number of
testable hypotheses from which we can generalize scientific
mechanisms~\citep{Efron2008}.  Occam's Razor motivates the $\ell_0$
penalty; conversely, estimating a contribution from every feature
contradicts a simple explanation and, downstream, produces a more
complex hypothesis to experimentally validate and to generalize across
correlated scientific samples.  The approach of model selection using
some approximation to the $\ell_0$ penalty has other downstream
benefits as well.  When the inclusion of a specific feature is modeled
explicitly, so that the posterior probability of inclusion is
estimated, we decouple the estimation of the effect size and the
inclusion of the feature~\citep{Bottolo2011}.
This has the effect of removing the \emph{Zero Assumption} (ZA) from
the statistical test, which is the assumption that coefficients with
effects near zero are null associations and should be
excluded~\citep{Efron2008}; the ZA is often false for a specific
application, as truly alternative associations often have effect sizes
near zero. Separately modeling signals and noise using a hierarchical
model leads to more natural tests for association directly on the
inclusion variable, and improves statistical power to detect
associated predictors with small effects~\citep{Polson2010}.

We examine the development of a Bayesian structured sparse prior
within a linear regression framework.  Let~${\by\in\reals^n}$ be the
continuous scalar responses for~$n$ samples.  Let~$\bX$ be
an~${n\times p}$ matrix of predictors.  We will use a linear
regression model with independent Gaussian noise:
\begin{align*}
  \by\given\bX,\bbeta,\nu &\sim
  \distNorm(\bX\bbeta, \nu^{-1}\eye_n),
\end{align*}
where~${\bbeta\in\reals^p}$ is the vector of coefficients,~${\nu >0}$ is the precision of the residual, and~$\eye_n$ is the~${n\times n}$ identity matrix.
The coefficients~$\bbeta$ are often referred to as \emph{effect sizes}, because each~$\beta_j$ captures the slope of the linear effect of predictor $j$ on the continuous response~\citep{Kendziorski2006,Stephens2009}. 

\subsection{Bayesian approaches to sparsity}
\label{sec:bayessparse}

Bayesian sparsity uses a prior distribution for model selection to
encourage a model to incorporate as few features as possible.  A
sparse prior on regression coefficients creates an \emph{a priori}
preference for~$\bbeta$ to be nonzero for only a subset of the
predictors.  In the absence of a detectable effect on the conditional
probability of~$\by$, a sparsity-inducing prior will encourage the
$\beta_j$ coefficient corresponding to predictor $j$ to be $0$,
indicating no linear association between predictor $j$ and the
response, and excluding predictor $j$ from the model.
Parameter estimation in sparse Bayesian regression is performed by
examining the posterior distribution on the~$\beta_j$ coefficients.

The canonical Bayesian sparsity-inducing prior is the spike-and-slab
prior~\citep{Mitchell1988,George1993}.  This \emph{two-groups} prior
introduces sparsity into the model via latent binary variables for
each predictor that capture whether that predictor is modeled as noise
or signal.  That is, each dimension of~$\bbeta$ is taken to be a
mixture of a Dirac delta function at zero (a~\emph{spike}, which
assigns positive probability to the event that predictor~$j$ has
exactly zero effect on the response) and a continuous density function
on~$\reals$ (a~\emph{slab}, which regularizes the included predictor,
allowing a wide range of possible contributions).  The spike-and-slab
formulation is
\begin{align}
  \label{eqn:spike-and-slab}
  \beta_j \given \omega, \tau &\sim
  \omega\,\delta(\beta_j) + (1-\omega)\,\distNorm(\beta_j\given 0, \tau^2)\;,
\end{align}
where we are modeling the ``slab'' as a zero-mean Gaussian with
variance~$\tau^2$.  The mixing parameter~${\omega\in[0,1]}$ determines
the expected proportion of sparse (excluded) components.  This
spike-and-slab prior is an elegant and direct approach to sparsity, in
line with~$\ell_0$ regularization: it results in posterior hypotheses
that directly quantify whether or not a particular predictor is
relevant to the response.  These hypotheses are explicitly represented
as latent inclusion variables~${z_j = \{0,1\}}$, which indicate
whether predictor~$j$ is included or excluded from the model.
When~${z_j = 0}$, the corresponding $\beta_j$ is distributed according
to a point mass at zero and when~${z_j=1}$,~$\beta_j$ is marginally
Gaussian.  Thus, we may interpret the posterior probability of $z_j$
as the probability that predictor~$j$ is included in the model, while
the estimated effect size of predictor~$j$ (conditioned on the
event~${z_j=1}$) is modeled separately in $\beta_j$.  There are a
number of variations on this version of the spike-and-slab prior, all
of which, to our knowledge, include an indicator variable that
captures the inclusion of predictor
$j$~\citep{Smith1996,Ishwaran2005,Guan2011}.  This elegant
interpretation motivates the use of a two-groups prior for the
regression framework when a sparse solution is desirable.

One of the challenges of the two-groups prior is the difficulty of
managing the hypothesis space for the~$z_j$ variables, which has~$2^p$
discrete configurations.  The need for computational tractability has
catalyzed the burgeoning field of continuous Bayesian
sparsity-inducing priors; these priors encourage the removal of
features from the model, but in a computationally tractable one-group
framework~\citep{Polson2010}.  One continuous prior is the Laplacian,
or double exponential, prior, the Bayesian analog to the
Lasso~\citep{Tibshirani1996,Hans2009}.
Another approach, the Gaussian-inverse gamma prior, called the
\emph{automatic relevance determination} (ARD) prior in the machine
learning literature~\citep{Tipping2001}, induces sparsity via a scale
mixture of Gaussians with an inverse gamma mixing measure on
predictor-specific variance terms.  This class of scale mixtures
corresponds to the Student's t-distribution when integrating over the
predictor-specific variance parameters, and, when the
degrees-of-freedom parameter~${\nu=1}$, it reduces to the Cauchy
distribution; both of these have been suggested as approaches to
one-group Bayesian sparsity~\citep{Polson2010}.  Other more recent
continuous sparsity-inducing priors include the
\emph{horseshoe}~\citep{Carvalho2009,Carvalho2010}, the generalized
double Pareto~\citep{Armagan2011}, the three parameter
beta~\citep{Armagan2011b}, and the Dirichlet Laplace
prior~\citep{Bhattacharya2012}.  These continuous priors differentially
modulate shrinkage effects for large and small signals while avoiding
the computational challenges associated with discrete sparsity models.
In particular, each of these one-group continuous priors has
substantial density around zero, shrinking noise to zero, and heavy,
sub-exponential tails, allowing large signals to remain intact,
without explicitly parameterizing predictor
inclusion~\citep{Mohamed2011}.  For a review of Bayesian
sparsity-inducing priors, see~\citet{Polson2010}.

While one-group priors are appealing for computational reasons as
continuous relaxations of discrete model-selection problems, they
obfuscate the core statistical questions surrounding estimates and
tests of inclusion \citep{Richardson2010}.  The posterior distribution
of parameter~$\beta_j$ captures the marginal effect size of the
predictor~$j$ on the response, and, in the context of model selection,
a zero-valued $\beta_j$ excludes predictor~$j$ from the model.  Under
continuous one-group priors and practical likelihoods,
however,~${\beta_j=0}$ measure zero under the posterior.  That is,
none of the $\beta_j$ variables will be exactly zero with positive
probability.  In practice, often a global threshold on the estimated
effect size is defined to determine model inclusion based on the
estimated ${\hat \beta}_j$ variables.  Using these posterior
distributions to evaluate model inclusion is not statistically
well-motivated with finite samples: features with small effect sizes
may be excluded because the resulting bimodal distribution of
significant effect sizes explicitly excludes predictors with estimated
effects in a region around zero. We prefer to avoid this zero
assumption and instead use a two-groups prior.

\subsection{Structured sparsity}
\label{sec:structure}

One assumption of the generic Bayesian regression model is that the
predictors are uncorrelated.  In practical applications, this
assumption is frequently violated~\citep{Breiman2001}.  The problem of
correlated or structured predictors has been studied extensively in
the classic statistical
literature~\citep{Jacob2009,Liu2008,Chen2012,Jenatton2011,Kim2009}, and
a number of methods that explicitly represent structure in the
predictors have been introduced for regression.  Group
Lasso~\citep{Yuan2006} uses a given disjoint group structure and
jointly penalizes the predictors group-wise using a Euclidean norm.
This model induces sparsity at the group level: the penalty will
either remove all features within a group or impose an $\ell_2$
penalty uniformly across features within a group.  This creates a
\emph{dense-within-groups} structure, as groups of predictors included
in the model will not be encouraged to have zero coefficients.  The
sparse group Lasso~\citep{Friedman2010} extended this idea by including
an $\ell_1$ penalty on the included groups' coefficients, creating a
\emph{sparse-within-groups} structure.  A Bayesian group Lasso model
has also been developed~\citep{Kyung2010}, where sparsity is encouraged
with a normal-gamma prior on the regression coefficients, and the
group structure is encouraged by sharing the gamma-distributed
variance parameter of the sparsity-inducing prior within a group.
This prior will also have a dense-within-groups structure, as all
included coefficients within a group will have a normal-gamma prior
with shared parameters, which will not induce zeros within included
groups.

Structured sparsity has proven to be useful in a wide variety of
practical applications such as image denoising, topic modeling, and
energy disaggregation \citep{kolter2010a,Jenatton2011b}. Despite its
utility in applied statistics, few proposals have been made for
Bayesian structured sparsity models; exceptions
include~\citet{Kyung2010}~and~\citet{Bottolo2011}.  This area of
research is ripe for innovation, as the Bayesian paradigm allows us to
incorporate structure naturally for heterogeneous data in a
hierarchical model to improve task performance.  Applied Bayesian
statisticians in particular may find that a Bayesian structured sparse
framework needs little tailoring to be customized to a specific
application other than a careful definition of the domain-specific
structure on the predictors, in contrast to the practical realities in
the broader literature~\citep{Kim2009,Jenatton2011b}.

In a Bayesian context, it is natural to impose structure through the
prior probability of the sparse parameters, sharing local shrinkage
priors between similar predictors~\citep{Kyung2010}.  Such hierarchical
models are straightforward to describe, and the Bayesian formalism
allows flexibility in the semantics of the structural representation.
For example, a common theme among many structured sparsity methods,
including group Lasso, Bayesian group Lasso, and tree
Lasso~\citep{Kim2009a}, is that the structure among the predictors is
defined as a discrete partition of feature space.  This disjoint
encoding of structure is not always possible, however, and many
applications require a more general notion of similarity between
predictors.  A flexible Bayesian approach would enable
application-driven ``soft'' measures of inclusion-relevant
similarity. This representational flexibility comes at a computational
cost, however, and such structured sparsity models must be designed
with considerations for the trade-off between the complexity of the
representation and inferential tractability.

In this paper, we describe a flexible Bayesian model for relaxed
two-groups sparse regression that includes a positive definite matrix
representing an arbitrary, continuous measure of similarity between
all pairs of predictors.  Despite considerable work on structured
sparse models in the classical framework, we are not aware of
substantial prior work on Bayesian structured sparse methods beyond
those mentioned above. We ground our modeling approach with a specific
motivating domain: the problem of associating genetic variants with
quantitative traits, which we describe in the proceeding section.  In
addition to our general-purpose model, we describe a Markov chain
Monte Carlo sampler for parameter estimation that enables the model to
be applied to large numbers of predictors by exploiting the structure
of the predictors to improve mixing.  We examine the empirical
performance of our approach by applying our model to simulated data
based on the motivating example of identifying genetic variants
associated with a quantitative trait.  Finally, we describe the
application of our model to identify genetic variants that regulate
gene expression levels across $40$ million genetic variants and more
than $17,000$ genes, and we compare results from our model to results
from a univariate approach to association mapping.

\section{Motivating application: associating genetic variants with quantitative traits}
\label{sec:ah}

\begin{figure}
  \centering 
  \hfill%
  \subfloat[SNP not associated with trait]{%
    \includegraphics[width=0.4\textwidth]{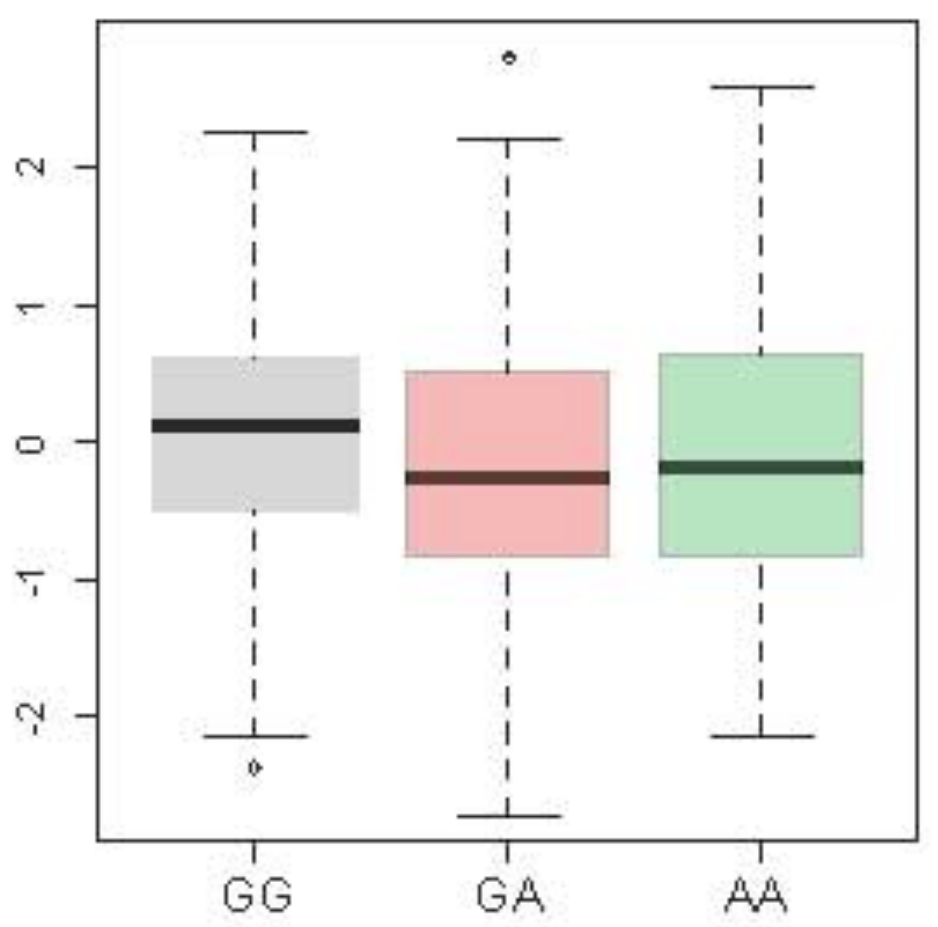}%
    \label{fig:no-association}
  }\hfill
  \subfloat[SNP associated with trait]{%
    \includegraphics[width=0.4\textwidth]{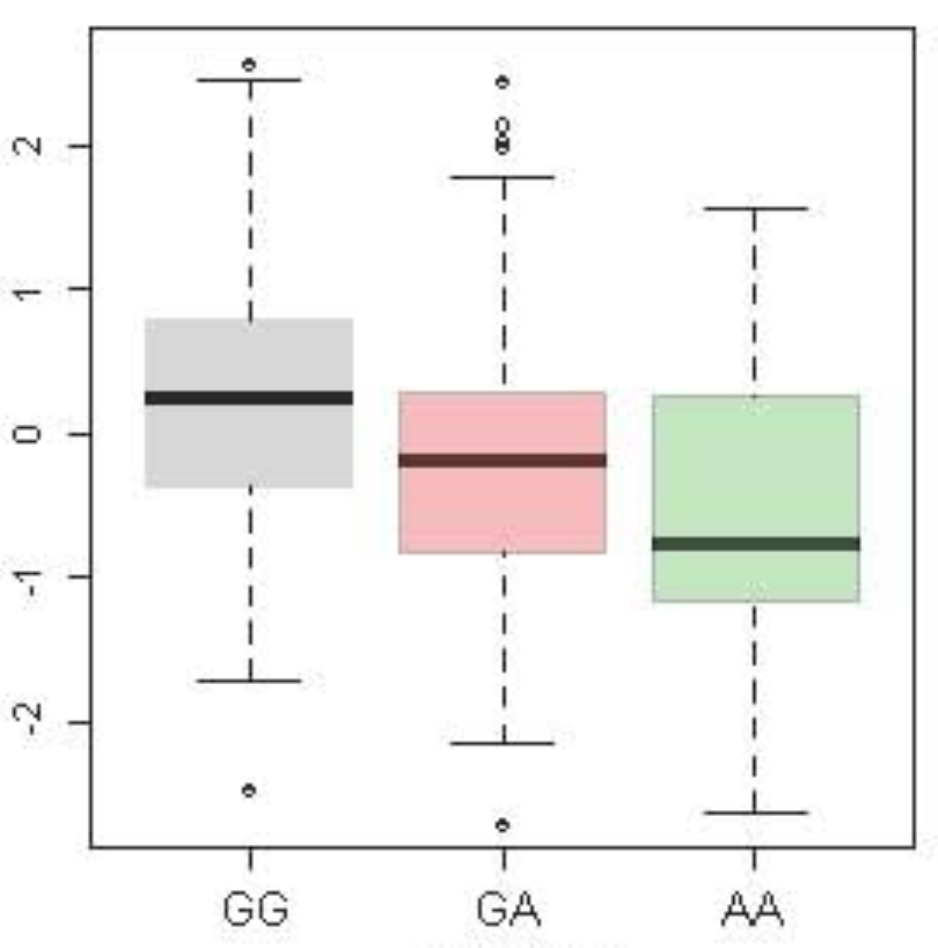}%
    \label{fig:association}
  }\hfill
  \caption{ \textbf{Examples of an SNP-trait pair with no evidence of
      association, and a SNP-trait pair with evidence of association.}
    For both, x-axis is a single SNP; y-axis is a single quantitative trait.
  }
  \label{fig:fake-association}
\end{figure}

Bayesian structured sparsity for linear regression is well-motivated
by the challenge of identifying genetic variants that are associated
with a quantitative trait, such as expression levels of a gene.  For
this problem, ${\by \in \reals^n}$ represents the quantitative trait
measurement across $n$ samples, and the predictors are single
nucleotide polymorphisms (\emph{SNPs}). We assume that each individual
has two copies of each nucleotide and that there are exactly two
possible \emph{alleles}, or variants, for each SNP. SNPs are encoded
as~${X_{i,j} \in \{0,1,2\}}$, which represent the number of copies of
the minor (or less frequent) allele for individual~${i \in 1,\dots,n}$
at one SNP~${j = 1,\dots,p}$.  The Bayesian testing problem is then to
determine which of the SNPs are associated with a given continuous
response, or quantitative trait; in other words, we wish to identify
each SNP $j$ where the true effect size on the trait, parameterized
through the $\beta_j$ coefficients, is non-zero
(Figure~\ref{fig:fake-association}).

The \emph{de facto} approach to identifying genetic associations
(called \emph{association mapping}) is to regress trait on SNPs
for~$n$ individuals, and then examine the magnitude of the estimated
linear coefficient~$\hat{\beta}$~\citep{Kendziorski2006}.  Testing for
association is performed by computing p-values or Bayes factors that
compare the likelihood given~${\beta = 0}$ (null hypothesis, no
effect) versus~${\beta \neq 0}$ (alternative hypothesis,
effect)~\citep{Stephens2009}.  Modeling and testing is often performed
in a univariate way, i.e., for a single trait and $p$ distinct SNPs,
we will have $p$ linear models and $p$ separate tests, one for each
SNP.  Typically, the number of individuals, $n$, is in the range of
$100$~--~$10,000$, and the number of SNPs, $p$, is in the range
$10,000$~--~$40,000,000$. The additional caveat is, in current
studies, there are often a large number of quantitative traits, as in
our application below.

A few approaches have combined these univariate models into a single
multivariate regression model, \emph{multi-SNP association mapping},
by including all $p$ SNPs as the predictors.  Then a sparsity-inducing
prior may be placed on the coefficient vector $\bbeta$ in order to
regularize appropriately in this $p \gg n$ regression with more
predictors (SNPs) than samples.  A sparse prior also matches our
belief that a small subset of SNPs will have a measurable regulatory
effect on a quantitative trait.  Indeed, multi-SNP association mapping
is an elegant example of an application of sparse models where the
underlying signal is thought to be truly sparse, as opposed to the
data being collected in such a way as to produce sparse
signals~\citep{Tibshirani2014}.

Sparse multivariate regression controls the combinatorial explosion of
univariate approaches by assuming additive effects across SNPs.  When
nonadditive interactions among the SNPs are present---called
\emph{epistatic effects}---these additive models are useful
approximations~\citep{Storey2005}, as current methods to detect
epistatic effects are infeasible due to a lack of statistical power.
Many multi-SNP association studies use greedy approaches to sparse
regression.  Forward stepwise regression
(FSR)~\citep{Brown2013,Stranger2012} is one greedy approach, with
stopping criteria defined by model scores such as AIC or
BIC~\citep{Schwarz1978}. The model starts with zero included SNPs, and
a SNP is included in the model if the model selection score improves
maximally with respect all other excluded SNPs; the algorithm
iterates, including a single SNP on each iteration, until none of the
excluded SNPs improves the current score.  Conditional analyses have
been proposed~\citep{LangoAllen2010,Yang2012,Yang2010} that identify
the most significant QTL association, and search for additional
single or pairwise associations conditional on the associations
identified thus far.  Penalized regression, specifically Lasso, has
also been used for multi-SNP association
mapping~\citep{Hoggart2008,Wu2009}.

Other model-based methods for detecting genetic associations in the
additive setting have used a combination of sparse regression and
model averaging~\citep{Brown2002}. Model averaging protects against the
substantial type I and type II errors that result from a non-robust
point estimate of the independently associated SNPs.  For example, if
two SNPs are perfectly correlated, or in perfect LD, and only one is a
causal SNP, a sparse method applied to the sample may select either
one of these SNPs with similar frequency.  In this case, there is
insufficient information from which to determine the causal variant,
and model averaging protects us from making the wrong choice.  A
recent approach used the Lasso, or $\ell_1$ penalized regression,
along with model averaging~\citep{Valdar2012} based on ideas from
stability selection~\citep{Meinshausen2010}.  Bayesian model averaging
has not often been applied to this problem for a number of reasons,
one of which is that the LD structure of the SNPs slows down the
mixing rate of sampling methods considerably.  Exceptions include the
MISA model~\citep{Wilson2010}, which uses Bayesian model averaging to
address the correlation in SNPs and also tests for association
directly on the posterior probability of inclusion.  A recent Bayesian
multivariate approach not only included $p$ SNPs but also $q$
traits, developing a multi-trait regression model with a matrix
response~\citep{Bottolo2011}.  This model used a sparse hierarchical
prior on the coefficients and clever methods for parameter estimation
in this high-dimensional space.  Other Bayesian methods have used
approximations to the spike-and-slab prior~\citep{Guan2011}, but not
included structure on the predictors or model selection.

\begin{figure}[t]
  \centering 
  \includegraphics[width=\textwidth]{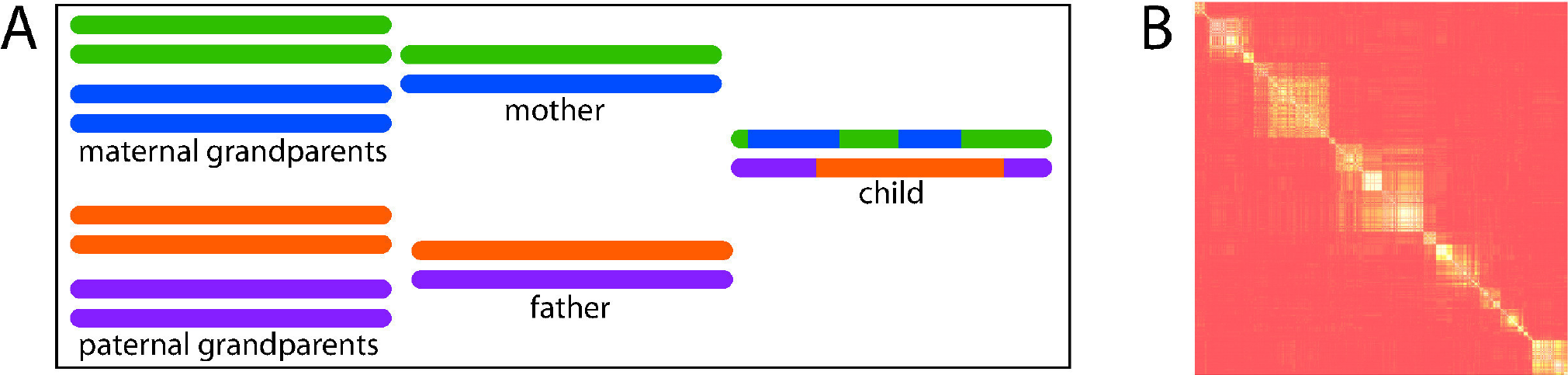} 
  \caption{ \textbf{Recombination and linkage disequilibrium in the
      human genome.}  \textbf{(A)} A single chromosome from each
    grandparent recombines in the parents;
    neighboring loci in the child chromosome will have
    identical grandparent of origin, except across sites of
    recombination.  \textbf{(B)} Heatmap of the absolute value
    Pearson's correlation among a thousand SNPs in a chromosomal
    region showing the block-like LD-structure in the genome.  }
  \label{fig:LD-structure}
\end{figure}

In this application there is substantial structure among the
predictors.  \emph{Linkage disequilibrium} (LD) refers to the
non-random assortment of genetic variants.  When offspring inherit one
complete set of chromosomes from each of their parents, each pair of
chromosomes that a parent inherited from his parents \emph{recombines}
in a handful of positions, so that a child receives a combination of
her grandparents' chromosomes for autosomal, or non sex-linked,
chromosomes (Figure~\ref{fig:LD-structure}). But because of the
relative infrequency of a recombination event, neighboring sites on a
child's chromosome are likely to be inherited together from the same
grandparent (Figure~\ref{fig:LD-structure})~\citep{Fledel2011}.  Local,
or \emph{background}, LD tends to result in block-like correlation
structure among SNPs on a chromosome
(Figure~\ref{fig:LD-structure})~\citep{Gabriel2002}.  The correlated
groups of SNPs are neither well-defined nor mutually exclusive, and
these correlations may exist across long genomic
distances~\citep{HapMap2005}.

In order to be useful in generating hypotheses for downstream
experimental validation and for use in clinical research, the goal of
association mapping---and \emph{fine mapping} in particular---is to
identify the \emph{causal SNP} for a given trait, or the SNP that, if
modified, would affect a change in that trait through biological
machinery.  The correlation structure between SNPs, however, means
that the identity of the causal SNP is uncertain within the set of
well correlated SNPs with a similar association significance.  Current
practice with univariate and hierarchical approaches select the SNP
with the greatest significance, but the assumption of exactly zero or
one associated SNPs does not match our understanding of genomic
regulation~\citep{Stranger2012,Wood2011}.  This assumption dramatically
oversimplifies the solution to the point of not producing robust,
interpretable
results~\citep{Maranville2011,Guan2011,Mangravite2013}. The biological
scenario where multiple genetic loci affect a trait through
independent mechanisms is called \emph{allelic
  heterogeneity}~\citep{Wood2011}. We take a sparse multivariate
approach to test for all independently associated SNPs.

A further difficulty of this application is that, because we observe a
subset of SNPs in our data, the causal SNP may be missing.  \emph{Tag
  SNPs} are correlated with a causal SNP and may confound the results:
if there are multiple tag SNPs for an unobserved causal QTL, it is
possible that they split the effect of the causal QTL, appearing
independently associated with the trait and acting as surrogate
predictors in the association model.

\section{A model for Bayesian structured sparsity}

We now describe our prior for two-group sparse regression using a
Gaussian field. We assume the data are~$n$
samples with $p$
predictors~$\{\bx_i,y_i\}^n_{i=1}$, ${\bx_i\in\reals^p}$,
and~${y_i\in\reals}$.  We will encode this response as an
$n$-vector~${\by\in\reals^n}$ and the predictors as a
matrix~${\bX\in\reals^{n\times p}}$.  The response variables are
conditionally independent, given the predictors and three parameters:
\begin{align}
  \by\given\bX,\bbeta,\beta_0,\nu &\sim
  \distNorm(\offset\bone_n + \bX\bbeta, \nu^{-1}\eye_n),
\end{align}
where we have separated out the offset~${\offset\in\reals}$,~$\bone_n$ is a length-$n$ column vector of ones,~${\bbeta\in\reals^p}$ is the vector of weights, and~${\nu>0}$ is the residual precision.
We place a (conjugate) gamma prior on the residual precision:
\begin{align}
  \nu &\sim \distGam(a_{\nu}, b_{\nu}).
\end{align}
For concreteness, we assume the following parameterization of the gamma distribution:
\begin{align}
  p(x\given a, b) &= \frac{b^a}{\Gamma(a)}x^{a-1}\exp\{-bx\}.
\end{align}
We place a zero-mean Gaussian prior on the offset:
\begin{align}
  \offset &\sim \distNorm(0, (\lambda\nu)^{-1})\;.
\end{align}

It is useful to recall that the Dirac delta function can be
interpreted as the limit of a zero-mean Gaussian distribution as the
variance goes to zero.  With this in mind, when conditioning on the
latent spike-and-slab inclusion variables~$z_j$ (where~${z_j=0}$
indicates exclusion through a ``spike'' and~${\beta_j=0}$,
and~${z_j=1}$ indicates inclusion, and the associated~$\beta_j$ is
drawn from the ``slab''), we form a degenerate diagonal covariance
matrix~$\bGamma$, where~${\Gamma_{j,j}=z_j}$, and write a
single~$p$-dimensional Gaussian prior that captures both included and
excluded predictors:
\begin{align}
\label{eqn:beta-prior}
\bbeta \given \nu, \lambda, \bGamma &\sim \distNorm(\bzero, (\nu\lambda)^{-1}\bGamma)\,.
\end{align}
Here~$\lambda$ is an inverse squared global scale parameter for the regression weights, on which we place a gamma prior:
\begin{align}
\lambda &\sim \distGam(a_{\lambda}, b_{\lambda})\;.
\end{align}
We follow Jeffreys~\citep{Jeffreys1961,Polson2010} and scale the global shrinkage parameter $\lambda$ by the residual precision $\nu$.

\subsection{Structure from a Gaussian field}

We introduce structure into the sparsity pattern of~$\bbeta$ by replacing independent~$z_j$---which in Eq.~\ref{eqn:spike-and-slab} would be Bernoulli with ${P(z_j=0)=\omega}$---with a probit link~\citep{Bliss1935,Albert1993} driven by a latent Gaussian field
\begin{align}
\label{eqn:gp}
\bgamma &\sim \distNorm(\bzero, \bSigma)\;.
\end{align}
The diagonal~$\Gamma_{j,j}$ is then determined by whether~$\gamma_j$ exceeds a threshold~$\gamma_0$, i.e.,~${\Gamma_{j,j}=\indic(\gamma_j > \gamma_0)}$.
We assume that the positive definite covariance matrix~$\bSigma$ is known; this matrix is used to specify the dependence structure for inclusion.
We complete the hierarchical model by placing a Gaussian prior on the probit threshold~$\gamma_0$:
\begin{align}
\label{eqn:gam0-prior}
  \gamma_0 &\sim \distNorm(\mu_{\gamma}, v_{\gamma})\;.
\end{align}
The marginal prior probability of inclusion of predictor $j$ is
computed directly via
\begin{align}
P(\beta_j \neq 0) &= 1 - \Phi\left(\frac{\gamma_0}{\Sigma_{j,j}}\right)\;,
\end{align}
where~$\Phi(\cdot)$ is the cumulative distribution function of the
standard normal.  Under the often reasonable restriction
that~$\bSigma$ be a correlation matrix with ones on the diagonal, the
expected number of included predictors is
is~${p\cdot(1-\Phi(\gamma_0))}$.  Conditioned on data, the posterior
probability of $\Gamma_{j,j}$ is the posterior probability of
inclusion (PPI) for the $j^{th}$ predictor.

In Equation~(\ref{eqn:gp}), the covariance matrix $\bSigma$ may be the
identity matrix, in which case this reduces to an unstructured
Bayesian model of sparse regression of Eq.~\ref{eqn:spike-and-slab},
with~${\omega=\Phi(\gamma_0)}$.  This formulation also encompasses
Bayesian (disjoint) group sparse regression when $\bSigma$ has a block
diagonal structure with constant non-negative values within each
block.  However, the advantage of this construction is the ability to
include structure in the sparsity-inducing prior, and so $\bSigma$ can
be an arbitrary positive definite matrix.  Of particular interest is
the case where the matrix~$\bSigma$ captures the pairwise similarity
between the predictors, i.e., the Gram matrix of a Mercer kernel.  As
with structured sparse regression models in the group Lasso and
related literature, predictors that are similar will have correlated
priors on the model exclusion parameters through the covariance
matrix. This has the effect of smoothing the $\gamma_j$ parameters for
similar predictors using a Gaussian process (GP).  GP structure has
been considered before for probit regression
models~\citep{Girolami2006}, but neither appear to have been used to
induce (structured) sparsity.

\subsection{Example: Structured Matrices for SNPs}

The choice of kernel function requires some thought for any
application.  For the problem of association mapping, we considered a
number of possible Mercer kernels that reflect the similarity between
SNPs, including:

\begin{itemize}
\item \emph{covariance}: $\Sigma^{cov} = X^TX$ 
\item \emph{absolute value Pearson's correlation}: $\Sigma^{cor}_{j,k} = \left|\frac { (X_j - {\bar X})^T (X_k - {\bar X})} {\sqrt{(X_j - {\bar X})^T(X_j - {\bar X})} \sqrt{(X_k - {\bar X})^T(X_k - {\bar X})}}\right|$, where ${\bar X}$ is the empirical mean of each feature $X_j$ across $n$ samples.
\item \emph{mutual information}: $\Sigma^{mi}$, a positive definite kernel based on a generalization of a Fisher Kernel~\citep{Seeger2000}.
\item \emph{centimorgan}: $\Sigma^{cM}$, a quantification of genetic linkage between two SNPs (empirically derived, e.g.,~\citep{1000genomes}).
\end{itemize}

We found that the choice of kernel within this set made minimal
difference in the simulation results, and for simplicity we chose the
absolute value Pearson's correlation kernel for our results.  In
practice, we ensure that each of these matrices are positive definite
by including a small regularization term on the diagonal.

Kernel functions are application specific; an arbitrary Mercer kernel
may be designed for other applications. While our choice of kernel in
this application is effective, it is a semantic approximation to our
true definition of ``similar'' in the setting of association
mapping. In particular, using this kernel it is difficult to
discriminate independently functional, but correlated, SNPs from
similarly functional, but correlated, SNPs. In other words, when two
SNPs are well correlated, and we find that they both are associated
with a gene via a univariate model, then two scenarios are possible:
i) conditioning on one of the SNPs in the same univariate model, the
other SNP is no longer correlated with the gene (the SNPs are
\emph{similarly functional}), or ii) in this conditional framework,
the other SNP remains associated with the gene levels (the SNPs are
\emph{independently functional}). Ideally, our kernel would indicate
high similarity for correlated and similarly functional SNP pairs, but
low similarity for correlated and independently functional SNP
pairs. Current work is focused on using additional biological
information, such as co-localized cis-regulatory
elements~\citep{Brown2013} and evolutionary
signatures~\citep{Rasmussen2013}, to achieve this quantification of
similarity and improve association mapping. This framework would also
allow us to invert the problem to investigate, in an iterative way,
evolutionary and genomic signatures of functional SNPs through
evaluation of specific kernels in discovering these truly causal
SNPs. Our Bayesian framework enables data-driven inference of the
relevant similarity between predictors, referred to as \emph{adaptive
  basis functions}, in which the kernel function is parameterized and
those parameters are estimated via the inference process. While we did
not estimate covariance parameters here, we expect this will be useful
as the biological semantics of the kernels become more complex.

Another comment about the kernel function focuses on the possible
concern that the data are used twice: once to estimate the Gram matrix
and again during inference to estimate model parameters. Given copious
available genomes, we suggest that the Gram matrix for this
application be estimated using reference genomic data, ideally from
the same population as the study data. Using reference data to
estimate the Gram matrix is well motivated because, in general, we
expect QTLs to replicate across studies~\citep{Brown2013}. Thus,
although there are caveats, a causal SNP in the study sample is likely
to be causal in the reference sample, and, by this reasoning,
similarities between SNPs should be meaningfully transferable.

\section{Parameter estimation with MCMC}

In many applications, the central quantities of interest are the
marginal PPIs.  In the context of genetic association studies, PPIs
allow us to test for association of the SNP with a trait, which is the
essential parameter for finding biologically functional genetic
variants.  These can be estimated via Markov chain Monte Carlo (MCMC)
using posterior samples of~$\bgamma$ and~$\gamma_0$.  Of particular
interest is the estimation of the posterior inclusion while
marginalizing out the effect size captured by~$\bbeta$.  The
degenerate Gaussian form of Eq.~\ref{eqn:beta-prior} makes it possible
to perform this marginalization in closed form and view the posterior
on~$\bgamma$ and~$\gamma_0$ through a regression marginal likelihood:
\begin{align}
    p(\by \given \bX, \bgamma, 
    \gamma_0, \nu, \lambda) &=
    \int\int \distNorm(\by \given \beta_0\bone_n + \bX\bbeta, \nu^{-1}\eye_n)\,
    \distNorm(\bbeta\given\bzero,(\nu\lambda)^{-1}\bGamma)\,
    \distNorm(\beta_0\given 0, (\nu\lambda)^{-1})\,\mathrm{d}\bbeta\,\mathrm{d}\beta_0\notag\\
    &= \int\distNorm(\by \given\beta_0\bone_n, \nu^{-1}(\lambda^{-1}\bX\bGamma\bX^{\trans} + \eye_n))\,\distNorm(\beta_0\given 0, (\nu\lambda)^{-1})\,\mathrm{d}\beta_0\notag\\
    &= \distNorm(\by \given \bzero, \nu^{-1}(\lambda^{-1}(\bone_n\bone_n^{\trans} 
    + \bX\bGamma\bX^{\trans}) + \eye_n))
    \label{eqn:marg-like}
\end{align}
In the Markov chain simulation, we update each of the
parameters~$\bgamma$, $\gamma_0$, $\lambda$, and~$\nu$ in
turn~\citep{Bottolo2011}.

\paragraph{Updating~$\bgamma$}

We use elliptical slice sampling (ESS), a rejection-free Markov chain
Monte Carlo (MCMC), for defining and simulating transition operators
on~$\bgamma$~\citep{Murray2010}.  ESS samples efficiently and robustly
from latent Gaussian models when significant covariance structure is
imposed by the prior, as in Gaussian processes and the present
structured sparsity model.  The conditional density on~$\bgamma$ is
the product of the likelihood in Eq.~\ref{eqn:marg-like} and prior in
Eq.~\ref{eqn:gp}.  ESS generates random elliptical loci using the
Gaussian prior and then searches along these loci to find acceptable
points for slice sampling.  When the data are weakly informative and
the prior is strong, as is the case here, the elliptical loci
effectively capture the dependence between the variables and enable
faster mixing.  Here, using ESS for~$\bgamma$ enables us to avoid
directly sampling over the large discrete space of sparsity patterns
that makes unstructured spike-and-slab computationally challenging.
We also note that elliptical slice sampling requires no tuning
parameters, unlike alternative procedures such as Metropolis--Hastings
or Hamiltonian Monte Carlo, which may mix faster but are often
difficult to tune and make robust.

\paragraph{Updating $\gamma_0$ and $\lambda$}
The scalar~$\gamma_0$ specifies the probit threshold and, conditioned
on~$\bgamma$, it determines which entries on the diagonal of~$\bGamma$
are zero and which are one.  The scalar parameter~$\lambda$ determines
the scale of the ``slab'' portion of the weight prior.  Neither of
these conditional densities has a simple closed form, but they can
each be efficiently sampled using the exponential-expansion slice
sampling algorithm described in \citep{Neal2003}.

\paragraph{Updating $\nu$}
The scalar~$\nu$ determines the precision of the residual Gaussian
noise of the response variables.  With the choice of a conjugate gamma
prior distribution, the conditional posterior is also gamma:

\begin{align}
p(\nu \given \by, \bX, \bGamma, \lambda) &\propto
\distNorm(\by\given\bzero, \nu^{-1}(\lambda^{-1}(\bone_n\bone_n^{\trans} + \bX\bGamma\bX^{\trans}) + \eye_n))\,\distGam(\nu\given a_\nu, b_\nu)\\
&= \distGam(\nu \given a^{(n)}_\nu, b^{(n)}_\nu)\\
a^{(n)}_\nu &= a_\nu + \frac{N}{2}\\
b^{(n)}_\nu &= b_\nu + \frac{1}{2}\by^{\trans}(\lambda^{-1}(\bone_n\bone_n^{\trans} + \bX\bGamma\bX^{\trans}) + \eye_n)^{-1}\by\,.
\end{align}

\section{Results}

To evaluate our model, we first used simulated trait data with
existing SNP data to compare methods where the complexity of the
predictor relationships was real, but the truth was known. We compared
our model against a number of other methods for association mapping
using precision-recall curves. Then, using these same SNP data, we
performed genome-wide association mapping with $16,424$ quantitative
traits, and compared the results from our method with results from
univariate Bayesian association mapping. The genomic data, from the
HapMap phase 3 project~\citep{HapMap2010}, include $608$ individuals
that we imputed to give us $40$ million SNPs per individual; a
complete description of these data, including quality control and
pre-processing performed, can be found in~\citet{Stranger2012} and~\citet{Brown2013}.

\subsection{Simulation Results}

Using simulations, we evaluated the performance of our model in
analyzing realistically complex genetic relationships, but with known
ground truth.  We simulated association data based on quantitative
traits sampled from a linear model with real SNP data as the
predictors.  In particular, given publicly available SNP data, we
first generated effect sizes for a small subset of randomly chosen
SNPs.  Next we generated a quantitative trait based on weighted linear
combinations of those included SNPs.

\paragraph{Simulating data.}

From the HapMap data, we selected $200$ SNPs at random from all SNPs
in one genetic locus, where loci were chosen from regions where we
expect SNPs to be functional (see below). We then selected at random
$q \in [2,6]$ QTLs per trait chosen from these $200$ SNPs $Q$. We
generated weights $\bbeta$, representing the effect size, from a
$\beta_j \sim \distBeta(0.1, 0.1)$ distribution, and we rescaled
$\beta_j$ to lie between $(-1, 1)$. The rationale behind this
simulation was to allow for effect sizes that were closer to 1 or -1
than to zero; this scenario will favor the methods that make the zero
assumption over ours. After generating a quantitative trait from $y_i
\sim \distNorm(Q_i\bbeta, 1)$, we projected each trait to the
quantiles of a standard normal (i.e., \emph{quantile normalized}
them). We repeated this for $696$ arbitrarily chosen genetic loci. We
call these data \emph{Sim1}.

A second simulation, \emph{SimTag}, considers the possibility that a
causal QTL may not be included in the set of SNPs, as may be the case
with genotyping data. In this simulation, we quantified how often the
methods incorrectly identify a \emph{tag SNP}, or a SNP that is
correlated with the causal SNP, when the causal SNP is missing. In
order to generate these data, we used an identical procedure as for
generating quantitative traits in Sim1, except we selected between $2$
and $8$ eQTLs ($s$) and chose some subset of those $r$ to remove from
the matrix $Q$ from a set of $200+r$ randomly chosen cis-SNPs,
where~${s-r>0}$. The subsequent steps---generating a trait using $s$
eQTLs and quantile normalizing the trait across individuals---are
identical to the steps in Sim1.

\paragraph{Methods for comparison}
Our evaluations compared results from a number of different methods,
some of which have been used for association mapping and others which
are common methods for model selection. We ran Lasso regression fitted
using Least Angle Regression (LARS), and identified the penalty term
by selecting the point in the LARS path with the smallest Bayesian
information criterion (BIC)
score~\citep{Hoggart2008,Efron2004,Schwarz1978}. We ran forward
stepwise regression (FSR), using the BIC score to determine when to
stop adding predictors to the model~\citep{Brown2013}. We ran Bayesian
sparse regression with an ARD prior on the weights
(ARD)~\citep{Tipping2001}. We applied our model of Bayesian sparse
regression with ${\Sigma = \eye_p}$ (identity matrix for the
covariance of the Gaussian field) to study our scale mixture
representation of spike-and-slab regression without structure on the
predictors (BSR).  We used $500$ iterations of burn-in for ESS and
$1000$ iterations for collection. We also ran our model of Bayesian
structured sparse regression with $\Sigma^{cor}$ to model the
correlation structure of the predictors (BSSR) with the same number of
ESS iterations as BSR. For BSSR, we also report results without model
averaging by selecting the configuration of the inclusion parameters
with the best posterior probability during sampling (MAPS).

We evaluated the results of applying our method to these simulated
data using precision-recall (PR) curves. In our simulations, only the
causal SNPs, or the SNPs selected in the simulation with a non-zero
effect, were considered \emph{truly alternative}; nearby SNPs that
were well correlated (or perfectly correlated) with the causal SNP but
had no simulated effect on the trait were considered to be \emph{truly
  null} in order to compute the true positive rates (TPR) and false
discovery rates (FDR)~\citep{Storey2003}. Results for FSR, Lasso, and
ARD are determined by thresholding the estimated effect sizes
$\hat{\bbeta}$; results from BSR, BSSR, and MAPS are determined by
thresholding the estimated PPI.

\paragraph{Comparison on simulated data.}

\begin{figure}[t]
  \centering 
  \subfloat[Precision-recall curve for simulating eQTLs]{%
    \includegraphics[width=0.48\textwidth]{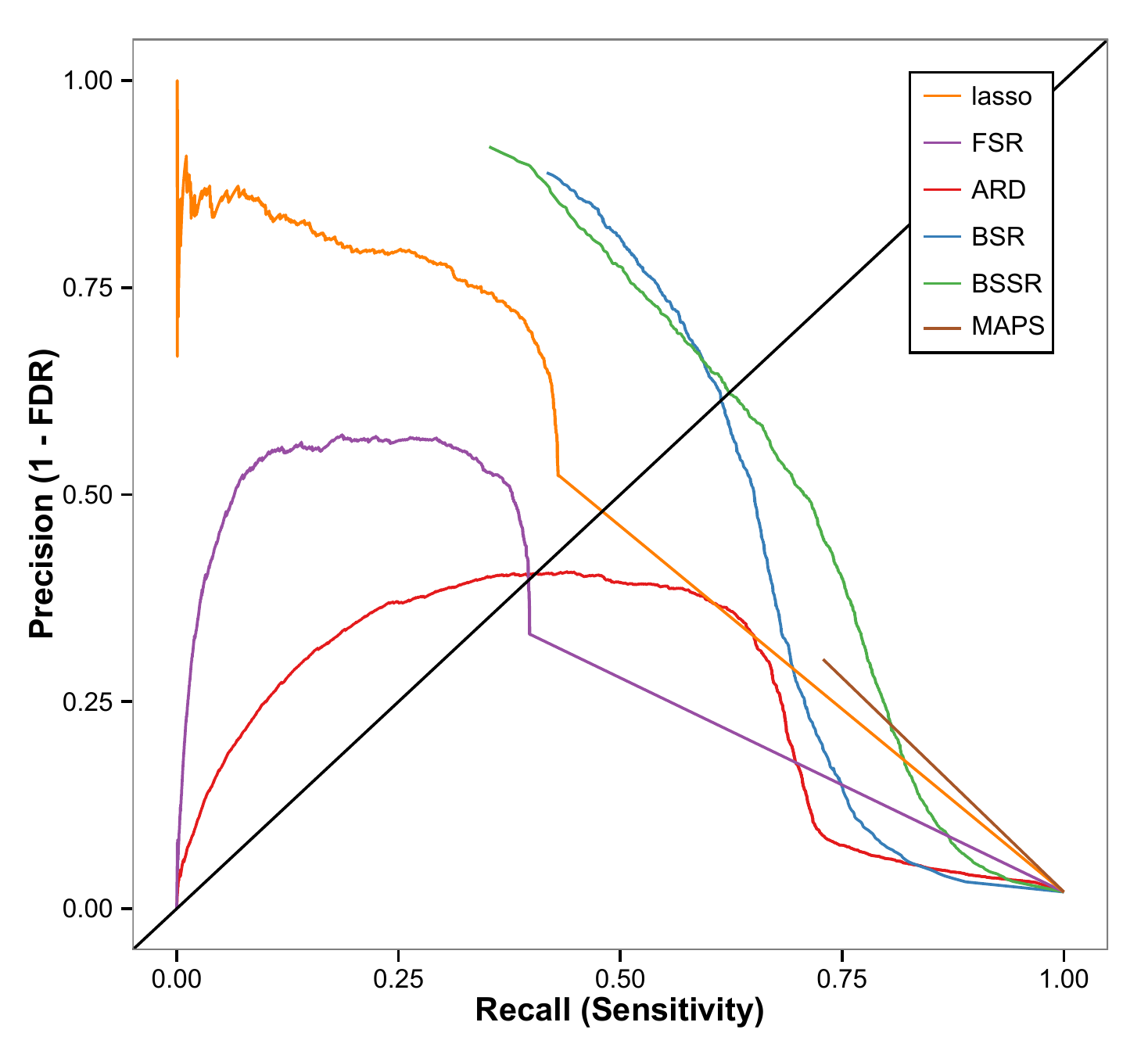}%
    \label{fig:results-sim1}
  }~
  \subfloat[Precision-recall curve for simulating tag SNPs]{%
    \includegraphics[width=0.48\textwidth]{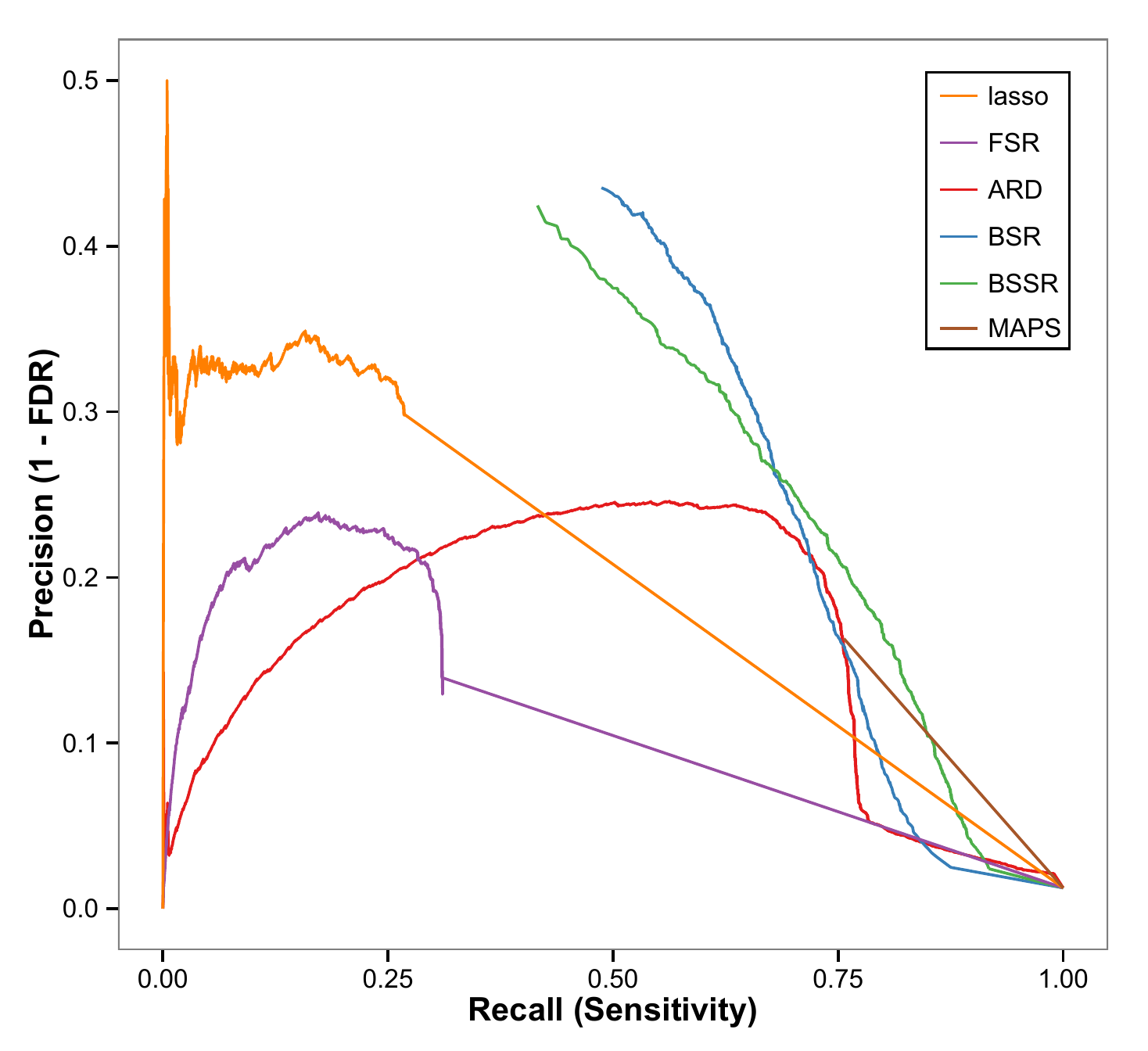}%
    \label{fig:results-sim2}
  }
  \caption{ \textbf{Precision-recall curves comparing five methods of
      association mapping using sparse regression methods on simulated
      data.} Precision-recall curves comparing the different methods
    along recall (true positive rate; x-axis) versus precision (1 -
    FDR; y-axis). The legend shows which curves correspond to which
    method.  
\textbf{(A)} Precision-recall curves for Sim1.
    \textbf{(B)} Precision-recall curves for SimTag.  }
  \label{fig:roc-sim}
\end{figure}

On Sim1, the comparative precision-recall curves show that BSSR
performs well across most levels of precision, particularly at high
precision, or, equivalently, low FDR (Figure~\ref{fig:roc-sim}). FSR
appears uniformly to give the poorest performance; at a threshold of
zero, the recall of FSR remains below 0.5. In the genomics community,
FSR is arguably the most common method to perform multi-SNP analyses
because of its intuitive simplicity~\citep{Stranger2012,Brown2013}. The
ARD prior also performs poorly. ARD has a fairly consistent
precision---under 0.4---across a large range of recall values; this
model does not have as much sparsity in the results as the other
methods.  Lasso performs better than FSR and ARD across all levels of
recall.  MAPS is a single point, because all PPI are either $0$ or
$1$, so all thresholds between these points have the same PR; the
recall is high, although the precision is lower than for BSR at that
same level of recall. For biological feasibility, we consider the best
comparison of the methods to be at a low FDR, as an FDR of $50\%$
indicates that there are equal numbers of true positive and false
positive biological hypotheses, decreasing their utility for expensive
downstream analysis or experimental validation.

Three observations about the relative performance of these methods
highlight the promise of our Bayesian approach to structured
sparsity. First, these simulations show the advantage of performing
association testing on the posterior probability of inclusion rather
than the absolute values of the regression coefficients themselves:
all of our Bayesian sparse models perform better in the PR curves than
any of the coefficient-based models (FSR, Lasso, ARD) at low
FDR.  The subtle but important distinction is that the
coefficient-based models make the Zero Assumption
(ZA)~\citep{Efron2008}, which is that the z-scores of the estimated
effect sizes near zero come from null associations, or, equivalently,
that p-values near one represent null associations. Indeed, this is a
central assumption in the q-value procedure to calculate
FDR~\citep{Storey2003}. Practically, by selecting a global threshold
for the $\beta_j$ variables, all associations with effect sizes of
smaller magnitude than the threshold are removed regardless of
evidence for association. 

PPI-based methods, however, do not make the ZA~\citep{Efron2008}; in
our two-groups model we have an explicit unimodal, zero-centered
distribution on the effect size of the included predictors.  Other
dense-within-groups formulations have been proposed, including
modeling the included predictors as a mixture of $K\ll p$ zero-mean
Gaussians or uniform distributions with their start or end points at
zero (Matthew Stephens, personal communication). We later suggest
inducing sparse-within-groups behavior by modeling the variance of
each of these Gaussian distributions separately.  This is the benefit
of using indicator variables in the global-local framework: each
coefficient $\beta_j$ is regularized globally (by parameter $\lambda$)
but included or excluded locally (by coefficient-specific parameters
$\gamma_j$), forcing weak signals to the null group, but rescuing low
effect size predictors through structure. A drawback to our approach
is that, for a similar FDR, the number of included variables in the
model is typically much higher than for methods that make the
ZA~\citep{Gelman2012}.

Second, there is evidence that adding structure to the covariates
improves the ability to perform association mapping: the precision
recall curve for BSSR (structure) shows distinct improvement over BSR
(no structure) except in a short region around 0.5 recall. Previous
work~\citep{Guan2011} has suggested that there is ``no good reason to
believe that the correlation structure of causal effects will follow
that of the SNPs.'' Our simulation results show that, while SNP
correlation is a gross proxy for the correlation structure of
independent causal effects, explicitly modeling SNP correlation does
improve our ability to perform association mapping.  

Third, model averaging appears to also confer an advantage
over the MAP configuration, which we observe when comparing the PRCs
of MAPS and BSSR results. It appears that model averaging protects
against type I errors and the non-robust point estimates of MAP
configurations, recapitulating previous
work~\citep{Valdar2012,Wilson2010}.

The comparative results for SimTag reflects qualitatively similar
performance, although the scale of the precision is on half of the PRC
figure for Sim1 (Figure~\ref{fig:roc-sim}). Interestingly, the
performance of MAPS did not decrease proportionally as much as the
other methods.

\begin{figure}[t]
  \centering
    \includegraphics[width=0.88\textwidth]{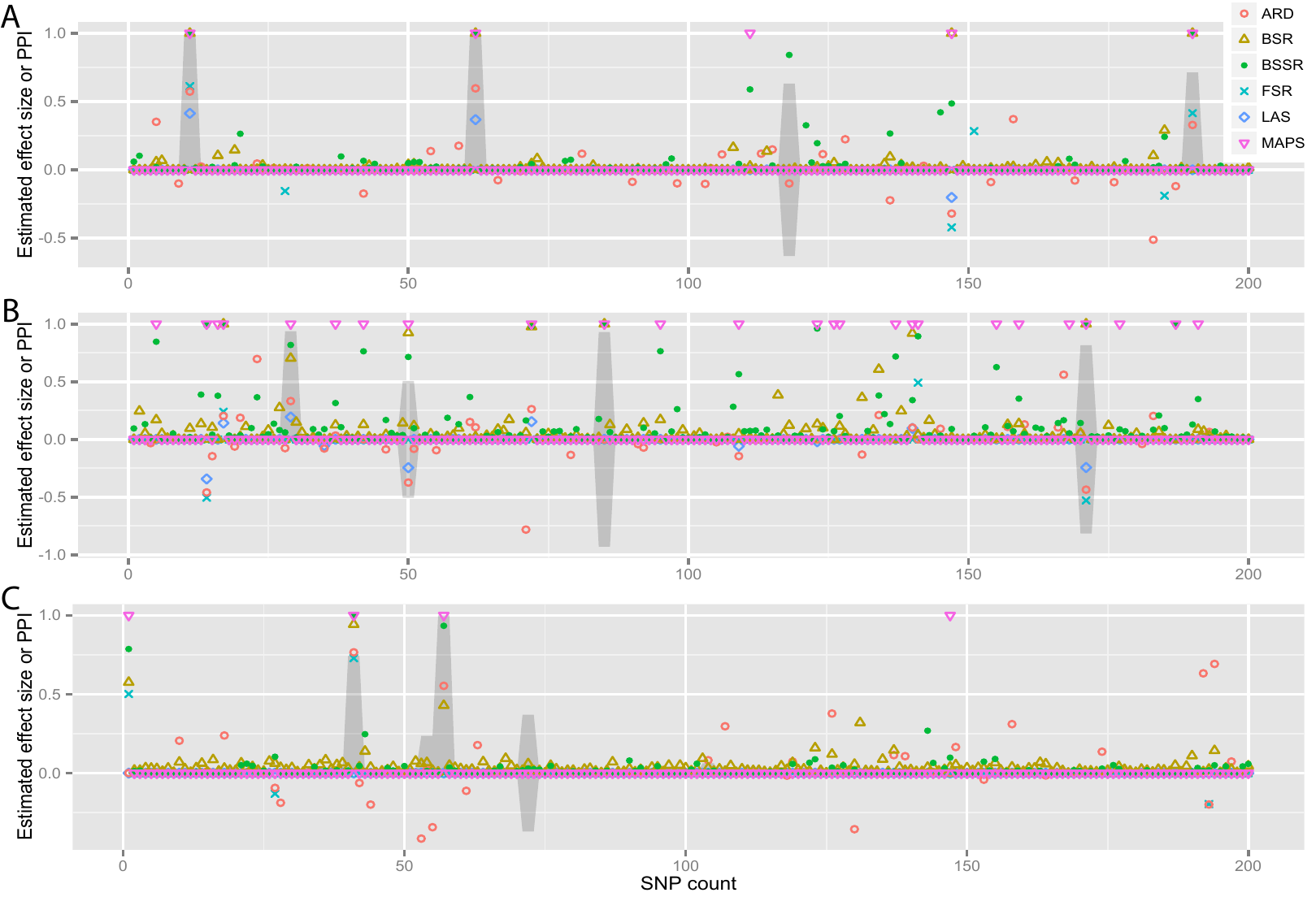}%
  \caption{ \textbf{Three examples of results from six methods on simulated quantitative trait data.} x-axis represents the SNP predictors, ordered along the chromosome; x-axis represents the estimated effect size (ARD, FSR, LAS) or the PPI (BSR, MAPS, BSSR). Truly alternative associations are shaded with the height of the shading representing their effect size; negative effect sizes are mirrored on the positive axis to highlight corresponding PPIs. Colors and shapes represent one of six methods shown in the legend; Red circles: Automatic relevance determination (ARD); Yellow triangles: Bayesian sparse regression (BSR); Green dots: Bayesian structured sparse regression (BSSR); Aqua X: Forward stepwise regression (FSR); Blue diamonds: Lasso regression (LAS); Pink triangles: MAP configuration for BSSR.
}
  \label{fig:sim-example}
\end{figure}

While precision-recall curves present aggregate results on these
simulated data, it is also instructive to study results from
individual simulations. We considered the results of six different
methods applied to three simulated traits with $200$ SNPs included in
Sim1 (Figure~\ref{fig:sim-example}). FSR appears to predict the truly
alternative predictors well across the three examples, but the number
of FP associations, and the substantial estimated effect size of those
FP associations, hurt the precision of the results. Echoing PR curve
findings, ARD results are dense, with many more non-zero effects
relative to results from other methods. Lasso appears to have a high
rate of false negatives, as reflected in its low sensitivity: Lasso
finds no associations in the third example. Furthermore, for Lasso the
FP predictors have estimates of the effect size equivalent to the
effect size predicted for the TP predictors (e.g., first and second
examples). BSR appears to identify the true positive values much of
the time in these three examples, but appears to be confounded by
correlation among the predictors, increasing the PPI of correlated
predictors, but often not to the level of the true positives. BSSR has
similar behavior because of its dense-within-groups behavior, and both
show greater sensitivity than other methods in these examples: BSSR
has two FNs with a conservative PPI threshold of $0.6$, and BSR has
three FNs. MAPS, the MAP configuration found during MCMC with BSSR
estimates each predictor inclusion as either $0$ or $1$; as a result,
there are more FPs than BSR and BSSR at similar sensitivity, but, in
these examples, only three FNs.

\subsection{Whole-genome cis-eQTLs}

To validate that our Bayesian structured sparse regression model could
be applied to whole-genome association mapping studies, we ran our
method with the $\Sigma^{cor}$ matrix on $16,242$ genes with gene
expression levels quantified using microarray data on the same $608$
HapMap phase 3 individuals sampled from $14$ distinct worldwide
regions~\citep{Stranger2012}. We compared the set of identified eQTLs
in these data with the eQTLs we identified using a univariate analysis
for Bayesian association mapping, SNPTEST~\citep{Marchini2007}.

Studies of expression quantitative trait loci (eQTLs) often limit the
number of tests for association by four orders of magnitude by
restricting the set of SNPs tested for any gene to the SNPs in
\emph{cis} with that gene, or local to that gene. Many studies have
suggested that, given our current limitations on sample size and the
small effects of each eQTL on gene expression levels, we should
restrict ourselves to identification of cis-eQTLs because local
effects tend to be larger and there are many fewer tests, resulting in
greater statistical
power~\citep{Mangravite2013,Stranger2012,Stranger2007,Morley2004}. In
this work, we restricted the SNPs tested for a given gene to those
that are located within $200$ Kb (or two hundred thousand bases) of
the transcription start site (TSS) or transcription end site (TES) of
a gene; thus, the size of the SNP window is $400$ Kb plus the size of
the gene.  For each gene, there is an average of $6,152$ cis-SNPs that
were tested, for a total of $100,039,937$ univariate gene-SNP tests.

\begin{figure}[t]
  \centering 
    \includegraphics[width=0.88\textwidth]{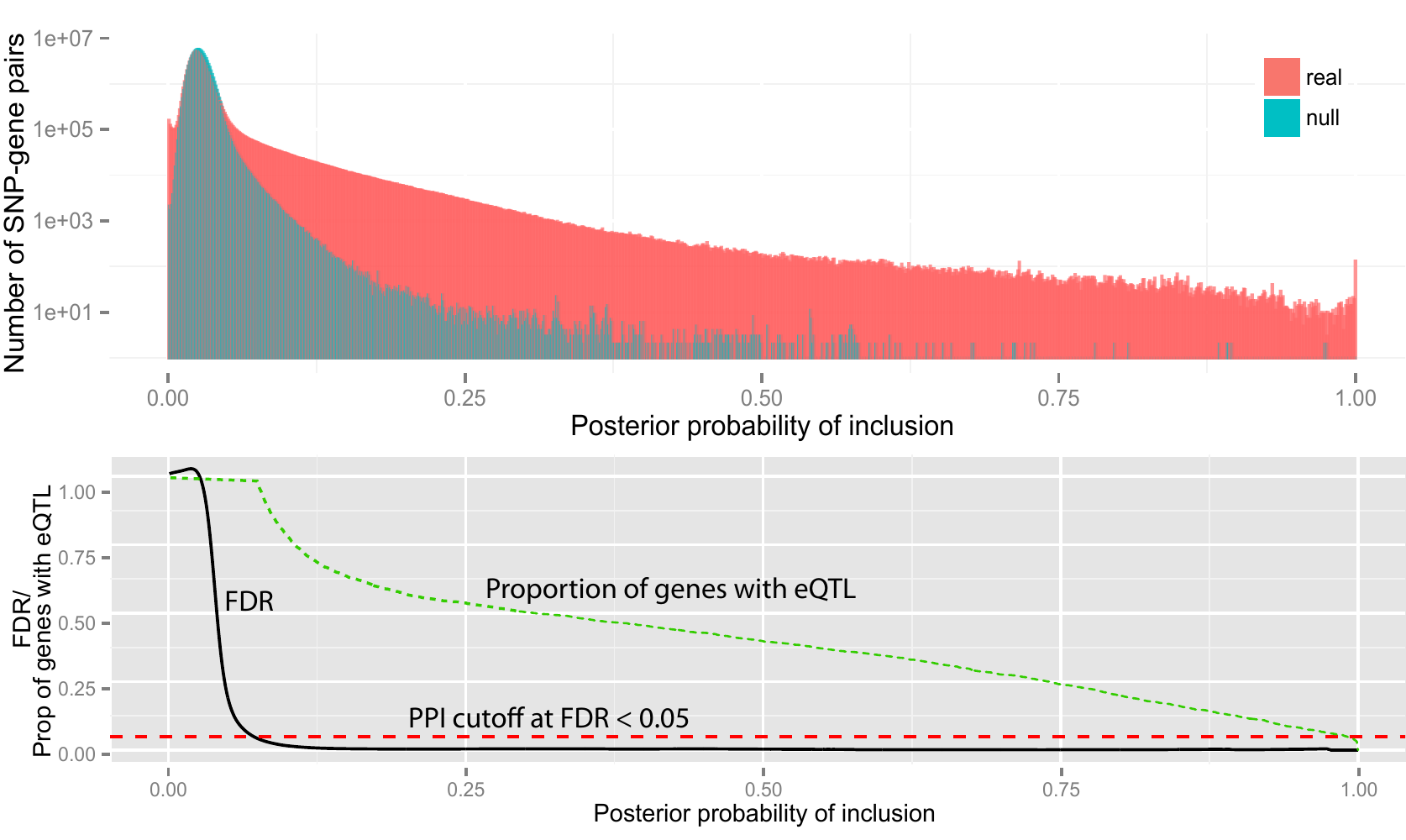}%
  \caption{ \textbf{Application of BSSR to real and permuted data.} The top histogram compares the number of SNP-gene pairs (y-axis; $\log_{10}$ scale) at each PPI (x-axis) for an identical number of tests in the real data versus the permuted (null) data; the bottom figure shows the estimated FDR from these tests, with the red dashed line indicating an FDR$=0.05$ and the green dotted line indicating the proportion of genes (out of the 2,589 total genes with eQTLs identified using BSSR) with one or more eQTLs at each PPI threshold. }
  \label{fig:results-ppi-fdr}
\end{figure}

We computed the FDR of a given threshold on the PPI from BSSR and on
the $\log_{10}$ Bayes factors (BF) from SNPTEST using a single
complete permutation of the data. Specifically, we permuted the sample
labels on the gene expression data matrix, and compute the inclusion
probabilities and $\log_{10}BF$s for every gene against every set of
cis-SNPs, under the assumption that these BFs will represent the same
number of tests under the null hypothesis of no association. Then,
across possible inclusion probabilities and $\log_{10}BF$ thresholds,
we computed FDR using the real and permuted results by:
\begin{eqnarray*}
\widehat{FDR}_{inc}(c_{ppi}) &=& \frac {\sum_{g = 1}^G \sum_{j=1}^p \indic(p(\Gamma_{j,j,g}) > c_{ppi})} {\sum_{g=1}^G \sum_{j=1}^p \indic(p(\Gamma_{j,j,g}^{perm}) > c_{ppi})} \\
\widehat{FDR}_{bf}(c_{bf}) &=& \frac {\sum_{g = 1}^G \sum_{j=1}^p \indic(\log_{10}BF_{j,g} > c_{bf})} {\sum_{g=1}^G \sum_{j=1}^p \indic(\log_{10}BF_{j,g}^{perm} > c_{bf})}, \\
\end{eqnarray*}
where $c_{ppi}$ and $c_{bf}$ are the PPI threshold the $\log_{10}BF$
threshold, respectively, and $p(\Gamma_{g,j,j})$,
$\log_{10}BF_{g,j,j}$ are the PPIs and the $\log_{10}BF$ of the
$g^{th}$ gene and the $j^{th}$ SNP, respectively, for $g =
\{1,\dots,G\}$ genes. The superscript $perm$ indicates that these
metrics were evaluated on the permuted data. We compared the number of
SNPs with $\log_{10}$BF with the observed and permuted data and
observed clear enrichment of large values
(Figure~\ref{fig:results-ppi-fdr}A). We selected the PPI threshold to
be $c_{ppi} = 0.072$ for FDR $\geq 5\%$.

We identified $2,940,533$ cis-eQTLs using our multi-SNP association
mapping (FDR $\leq 5\%$, PPI $\geq 0.073$), as compared to $169,460$
cis-eQTLs found using the univariate association mapping method (FDR
$\leq 5\%$, $\log_{10}BF \geq 1.092$). There were $2,589$ out of
$16,242$ genes with at least one eQTL for BSSR versus $5,065$ genes
with at least one eQTL for the univariate association mapping method;
of these, $1,463$ genes had significant eQTLs for both approaches
(FDR$\leq 5\%$). This order-of-magnitude increase in the number of
cis-eQTLs is not unexpected given that our approach explicitly removes
the zero assumption~\citep{Gelman2012}; but, accompanied with a
decrease in the total number of genes with at least one eQTL, suggests
that our method improves the precision of this approach when viewed in
a gene-by-gene way (Figure~\ref{fig:results-ppi-fdr}B). These results
suggest that the null hypothesis of no association may be
incorrect in a multi-SNP setting: the PPI of a truly null predictor
has less shrinkage toward zero when it is correlated with a truly
alternative predictor, increasing the average PPI of the truly null
predictors that are correlated with a truly alternative predictor. For
genes with no causal SNPs, this null hypothesis was appropriate.  We
suggest presenting results for this model in rank order by gene to
experimental biologists, indicating our confidence with the cis-eQTLs
with largest PPIs for follow up experimental validation.

\begin{figure}[t]
  \centering 
  \subfloat[Genes with more BSSR eQTLs than univariate eQTLs]{%
    \includegraphics[width=0.48\textwidth]{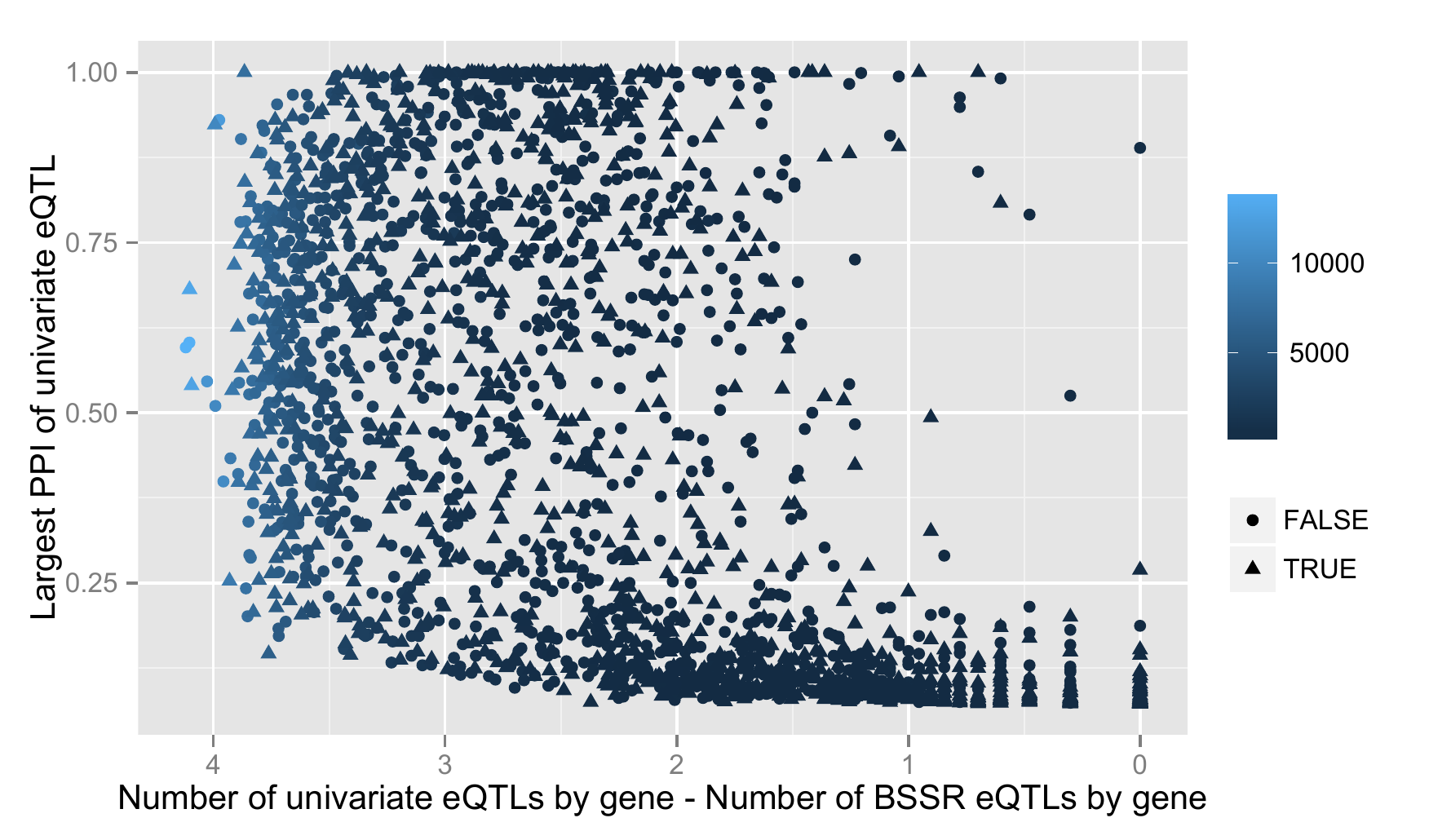}%
    \label{fig:bssr-vs-univ-bssr}
  }\hfill
  \subfloat[Genes with more univariate eQTLs than BSSR eQTLs]{%
    \includegraphics[width=0.48\textwidth]{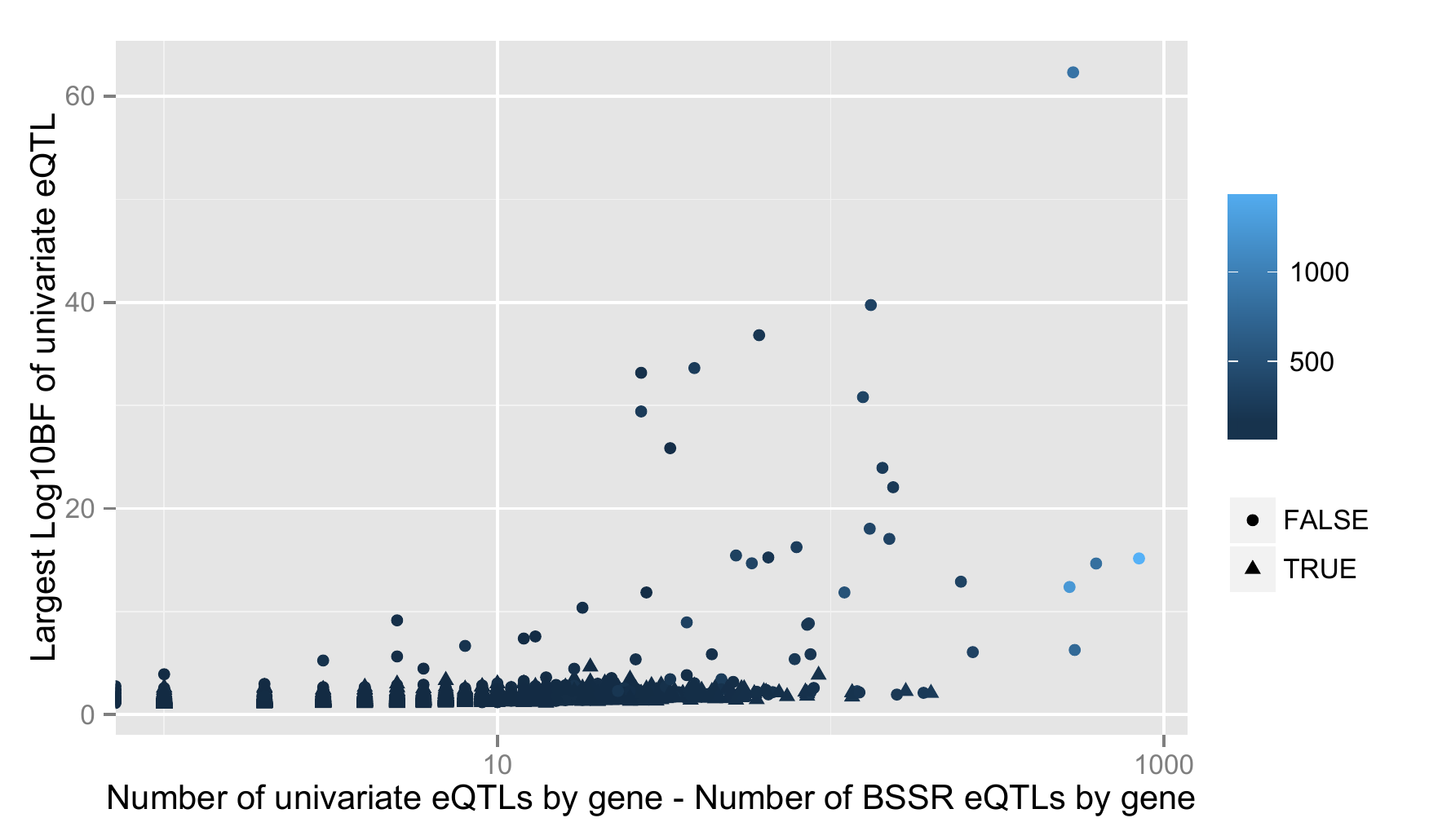}%
    \label{fig:bssr-vs-univ-univ}
  }
  \caption{ \textbf{Comparison of eQTLs identified by univariate and BSSR approaches by gene.} Each point represents a single gene for which at least one eQTL was identified using one of the two methods. X-axis ($\log_{10}$ scale) is the difference in the number of eQTLs per gene by the univariate approach and the BSSR approach. Triangles represent genes for which the approach with fewer eQTLs identified zero eQTLs for that gene; circles represent genes for which the approach with fewer eQTLs identified at least one eQTL for that gene.
\textbf{(A)} y-axis is the largest PPI for an eQTL within that gene.
    \textbf{(B)} y-axis is the largest $\log_{10}BF$ for an eQTL within that gene.  }
  \label{fig:diff-bssr-univ}
\end{figure}

It is instructive to compare the results from BSSR to those from the
univariate approach on a gene-by-gene basis. When we consider those
genes for which the univariate approach found more eQTLs than the BSSR
approach (Figure~\ref{fig:diff-bssr-univ}, right hand side) we notice
that, for all of the genes where BSSR identified zero eQTLs, the
univariate approach did not find any eQTLs with a $\log_{10}BF \geq
5$, indicating possible false positives that are systematically
eliminated using our model. In particular, there are $3,602$ genes
with zero BSSR-identified eQTLs, which only have a few
univariate-derived eQTLs each, all with low $\log_{10}BF$s; these
non-replicating associations across genes are candidates for false
positive associations. Genes with univariate-derived eQTLs with the
greatest statistical significance have at least one significant
BSSR-derived eQTL; fewer BSSR-derived eQTLs for that gene may indicate
weaker LD in this region. Conversely, we also considered the set of
genes for which there were more BSSR-derived eQTLs than
univariate-derived eQTLs (Figure~\ref{fig:diff-bssr-univ}, left-hand
side). While there are many fewer genes with zero univariate-derived
eQTLs in this set ($1,126$ genes), these genes often include SNPs with
large PPI. Echoing results from the simulations, this suggests that
BSSR approaches are able to identify eQTLs with smaller effect sizes
by exploiting small effect sizes across structured SNPs, resulting in
high recall. On the other hand, there also appears to be inflation of
eQTL signals in BSSR results, suggesting a sparse-within-groups model
may have improved precision on a SNP-by-SNP basis.

\begin{figure}[t]
  \centering 
    \includegraphics[width=0.88\textwidth]{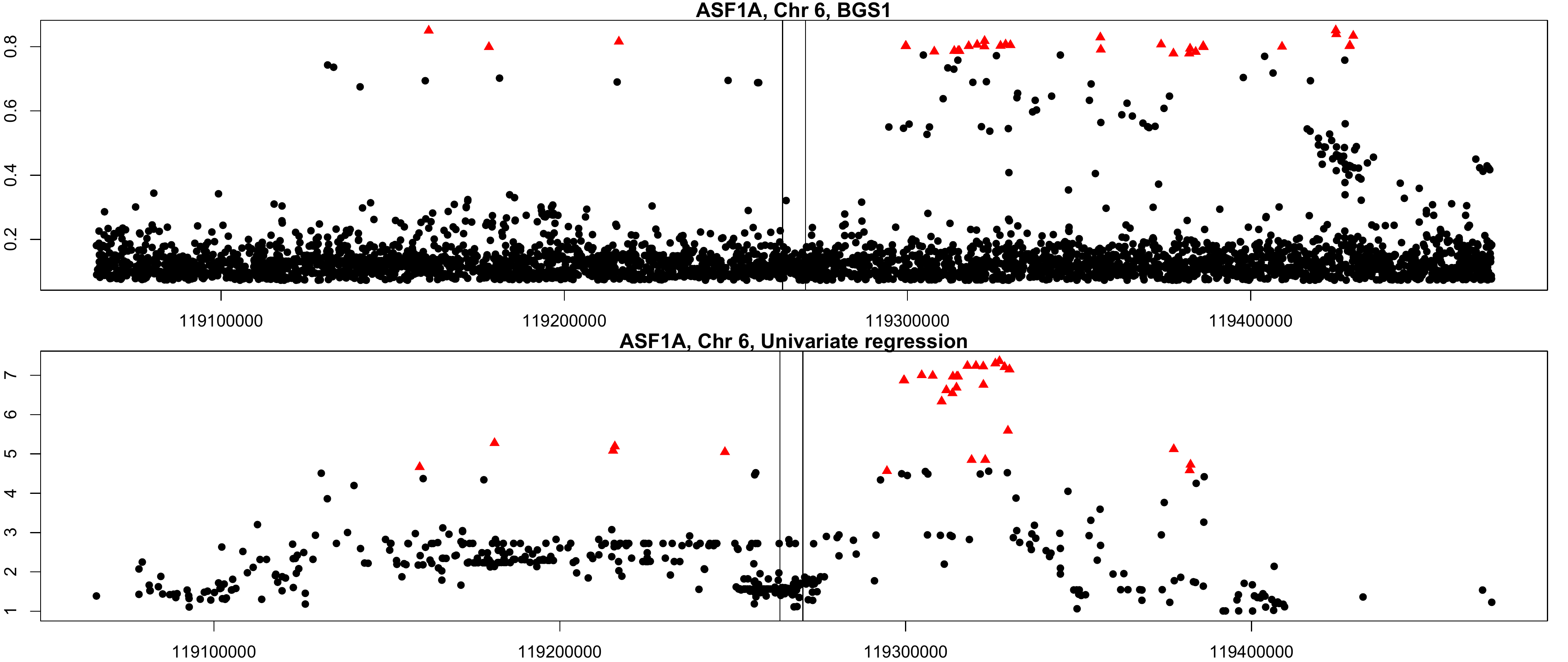}

    \includegraphics[width=0.88\textwidth]{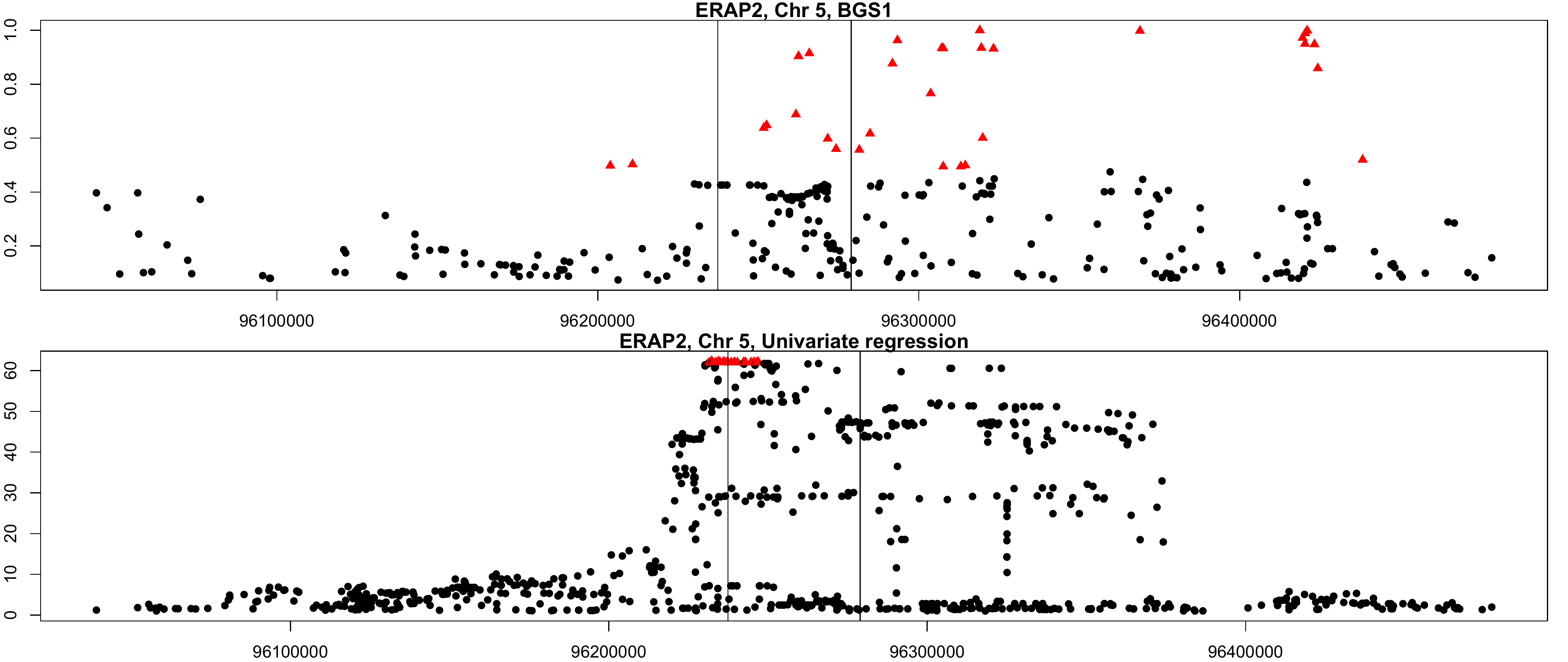}%
  \caption{ \textbf{Results from BSSR and univariate approaches on two genes: ASF1A and ERAP2.} The x-axis is the chromosomal position of each of the SNPs; the y-axis is the PPI (BSSR) or the $\log_{10}BF$ (univariate regression) of the SNP. Black circles represent statistically significant eQTLs (FDR$\leq 5\%$), red triangles represent the top $31$ eQTLs for this gene from the two approaches. Horizontal lines denote the transcription start and end site of the gene.
    }
    \label{fig:two_real_genes}
\end{figure}

We considered results from two specific genes
(Figure~\ref{fig:two_real_genes}). For \emph{ASF1A}, we find that
there are many more significant BSSR-identified eQTLs than
univariate-identified eQTLs. We suspect that most of the eQTLs below
$0.4$ are included because they are well correlated with a causal
SNP. In the case that the SNP is truly null, these eQTLs will be
removed by a sparse-within-groups model; in the case that multiple
causal SNPs are correlated, the sparse signal is effectively split
between well correlated SNP, reducing the marginal PPIs. From the
perspective of generating testable hypotheses, we highlight the $31$
most significant SNPs, and find their range almost identical to the
$31$ most significant SNPs in the univariate regression
analysis. Results from a second gene, \emph{ERAP2}, show the opposite:
the $31$ most significant SNPs from the univariate analysis do not
appear in the BSSR analysis, and the BSSR analysis highlights SNPs
across a much larger range of the genome. On the far right, BSSR
identifies clear signal that is poorly ranked in the univariate
analysis because of small effect size. SNPs with similar PPI and
$\log_{10}BF$ in both the BSSR and the univariate analysis
(identifiable as lines of points looking vaguely horizontal)
illustrate signal splitting tendencies (in the case of BSSR) and the
difficulties of finding the causal SNP amongst highly correlated
predictors (in the case of the univariate analysis).

\section{Discussion}

In this manuscript, we introduced a general Bayesian structured sparse
prior, encoding structure in the predictions via a Gaussian field,
using an arbitrary positive definite matrix. We described its
application in the context of regression to identify associations
between genetic variants and quantitative traits where there is
substantial structure in the genetic variants. We found that our prior
robustly identifies directly associated predictors, and includes a
natural statistical test for association. For recovering quantitative
trait associations, we found many more associations per trait, but,
across multivariate regression models, we found a smaller number of
responses with one or more associated predictors across all tests.

In the original paper on group sparsity, the method encourages
\emph{dense within groups} sparsity, where either every member of the
group was shrunk to zero or had minimal
regularization~\citep{Yuan2006}. However, there are good arguments in
favor of \emph{sparse within groups} sparsity, which shrinks groups to
zero together, but also encourages individual
sparsity~\citep{Friedman2010}. In practice this decision is tailored to
the application and the interpretation of the variables. The latter is
certainly more natural in the framework of correlated coefficients,
because we would like to select the smallest number of covariates that
explain the variation in the data rather than a dense set with
redundancies.  However, from the Bayesian perspective, the
sparse-within-groups model does not have a posterior mode that is
robust to sample bias, and, instead, Bayesian statisticians find the
dense-within-groups model more interpretable and
generalizable~\citep{Valdar2012}. Our formulation of Bayesian
structured sparsity is dense-within-groups; however there are
straightforward ways to tailor the model to achieve explicit
sparse-within-groups performance, in particular, to modify the scalar
global regularization term $\lambda$ to be local $p$-vector predictor
specific regularization. Because we are working within a multivariate
regression framework, if two predictors have a similar effect on the
response, the model will tend to select one for inclusion, and the
posterior probability of inclusion for the two predictors will split
the effect. This is where model averaging is helpful in encouraging
the choice of included predictors to be robust.

A number of other extensions are interesting to consider in light of
the application of these models to high-dimensional data. First, while
our sampler is efficient and produces an estimate of the
marginal PPI for each predictor, we would like to scale this to
perform analytical association mapping for genome-scale data sets,
which currently include approximately $40$ million SNPs and thousands
or tens of thousands of individuals~\citep{Jallow2009}. One option is
to window the entire genome into blocks of $10,000$ SNPs, and perform
association mapping within each of these windows in parallel. While
this is easy to do, it is somewhat unsatisfying. It is desirable to
consider other approaches, including multiscale methods and robust
adaptations of stochastic variational methods for $p \gg n$
applications. Second, the additive assumptions implicit in this model
are, with a few exceptions, made across this field of
research~\citep{Storey2005}, but ideally we would identify epistatic
effects, or non-additive interacting effects, between predictors.

While we have presented this structured sparse framework based on GP
probit regression, there are a large number of alternatives to this
specific choice of prior that may be explored while maintaining
tractability and expressiveness of this framework.

\section{Conclusions}

We present a general formulation for Bayesian structured sparsity that
includes information about covariate structure in a positive definite
matrix. Shrinkage is shared across inclusion variables for similar
covariates via a Gaussian field.  Applying this approach to regression
models for association mapping for quantitative traits, we find that
this method has a number of statistical and computational advantages
over current approaches. Furthermore, the arbitrary positive definite
matrix allows the model to be tailored to arbitrary applications using
domain-specific measures of similarity between predictors and that the
pairwise similarity may be arbitrarily complex. Computationally, these
methods are tractable for large studies and will be useful for many
applications of structured sparsity and model selection.  Accompanying Python code is available at \url{https://github.com/HIPS/BayesianStructuredSparsity}.

\subsection*{Acknowledgements}
BEE was funded by BEE was funded through NIH NHGRI R00-HG006265 and
NIH NIMH R01-MH101822. RPA was partially funded by DARPA Young Faculty
Award N66001-12-1-4219.

\bibliographystyle{plainnat}
\bibliography{bgsr}

\begin{thebibliography}{73}
\providecommand{\natexlab}[1]{#1}
\providecommand{\url}[1]{\texttt{#1}}
\expandafter\ifx\csname urlstyle\endcsname\relax
  \providecommand{\doi}[1]{doi: #1}\else
  \providecommand{\doi}{doi: \begingroup \urlstyle{rm}\Url}\fi

\bibitem[Albert and Chib(1993)]{Albert1993}
James~H. Albert and Siddhartha Chib.
\newblock {Bayesian analysis of binary and polychotomous response data}.
\newblock \emph{Journal of the American Statistical Association}, 88\penalty0
  (422):\penalty0 669--679, 1993.

\bibitem[Armagan et~al.(2011{\natexlab{a}})Armagan, Dunson, and
  Lee]{Armagan2011}
Artin Armagan, David Dunson, and Jaeyong Lee.
\newblock {Generalized double Pareto shrinkage}.
\newblock \emph{preprint arXiv:1104.0861}, 2011{\natexlab{a}}.

\bibitem[Armagan et~al.(2011{\natexlab{b}})Armagan, Dunson, and
  Clyde]{Armagan2011b}
Artin Armagan, DB~Dunson, and Merlise Clyde.
\newblock {Generalized beta mixtures of Gaussians}.
\newblock \emph{arXiv preprint arXiv:1107.4976}, pages 1--9,
  2011{\natexlab{b}}.

\bibitem[Bhattacharya et~al.(2012)Bhattacharya, Pati, Pillai, and
  Dunson]{Bhattacharya2012}
Anirban Bhattacharya, Debdeep Pati, Natesh~S. Pillai, and David~B. Dunson.
\newblock {Bayesian shrinkage}.
\newblock December 2012.

\bibitem[Bliss(1935)]{Bliss1935}
C.~I. Bliss.
\newblock The calculation of the dosage-mortality curve.
\newblock \emph{Annals of Applied Biology}, 22\penalty0 (1):\penalty0 134--167,
  February 1935.

\bibitem[Bottolo et~al.(2011)Bottolo, Petretto, Blankenberg, Cambien, Cook,
  Tiret, and Richardson]{Bottolo2011}
Leonardo Bottolo, Enrico Petretto, Stefan Blankenberg, Fran\c{c}ois Cambien,
  Stuart~a Cook, Laurence Tiret, and Sylvia Richardson.
\newblock {Bayesian detection of expression quantitative trait loci hot spots.}
\newblock \emph{Genetics}, 189\penalty0 (4):\penalty0 1449--59, December 2011.

\bibitem[Breiman(2001)]{Breiman2001}
Leo Breiman.
\newblock {Statistical modeling: The two cultures}.
\newblock \emph{Statistical Science}, 16\penalty0 (3):\penalty0 199--231, 2001.

\bibitem[Brown et~al.(2013)Brown, Mangravite, and Engelhardt]{Brown2013}
Christopher~D. Brown, Lara~M. Mangravite, and Barbara~E. Engelhardt.
\newblock {Integrative Modeling of eQTLs and Cis-Regulatory Elements Suggests
  Mechanisms Underlying Cell Type Specificity of eQTLs}.
\newblock \emph{PLoS Genetics}, 9\penalty0 (8), August 2013.

\bibitem[Brown et~al.(2002)Brown, Vannucci, and Fearn]{Brown2002}
P.~J. Brown, M.~Vannucci, and T.~Fearn.
\newblock {Bayes model averaging with selection of regressors}.
\newblock \emph{Journal of the Royal Statistical Society: Series B (Statistical
  Methodology)}, 64\penalty0 (3):\penalty0 519--536, August 2002.

\bibitem[Carvalho et~al.(2009)Carvalho, Polson, and Scott]{Carvalho2009}
Carlos~M. Carvalho, Nicholas~G. Polson, and James~G. Scott.
\newblock Handling sparsity via the horseshoe.
\newblock \emph{Journal of Machine Learning Research: Workshop and Conference
  Proceedings}, 5:\penalty0 73--80, 2009.

\bibitem[Carvalho et~al.(2010)Carvalho, Polson, and Scott]{Carvalho2010}
Carlos~M. Carvalho, Nicholas~G. Polson, and James~G. Scott.
\newblock The horseshoe estimator for sparse signals.
\newblock \emph{Biometrika}, 97\penalty0 (2):\penalty0 465--480, 2010.

\bibitem[Chen et~al.(2012)Chen, Lin, Kim, Carbonell, and Xing]{Chen2012}
Xi~Chen, Qihang Lin, Seyoung Kim, Jaime~G. Carbonell, and Eric~P. Xing.
\newblock {Smoothing proximal gradient method for general structured sparse
  regression}.
\newblock \emph{The Annals of Applied Statistics}, 6\penalty0 (2):\penalty0
  719--752, June 2012.

\bibitem[Consortium et~al.(2005)]{HapMap2005}
International~HapMap Consortium et~al.
\newblock A haplotype map of the human genome.
\newblock \emph{Nature}, 437\penalty0 (7063):\penalty0 1299--1320, 2005.

\bibitem[Consortium et~al.(2010)]{HapMap2010}
International HapMap~3 Consortium et~al.
\newblock Integrating common and rare genetic variation in diverse human
  populations.
\newblock \emph{Nature}, 467\penalty0 (7311):\penalty0 52--58, 2010.

\bibitem[Durbin et~al.(2010)Durbin, Altshuler, and Abecasis]{1000genomes}
Richard~M. Durbin, David~L. Altshuler, and Gon\c{c}alo R.~\emph{et al.}
  Abecasis.
\newblock {A map of human genome variation from population-scale sequencing}.
\newblock \emph{Nature}, 467\penalty0 (7319):\penalty0 1061--1073, October
  2010.

\bibitem[Efron(2008)]{Efron2008}
Bradley Efron.
\newblock {Microarrays, Empirical Bayes and the Two-Groups Model}.
\newblock \emph{Statistical Science}, 23\penalty0 (1):\penalty0 1--22, March
  2008.

\bibitem[Efron and Hastie(2004)]{Efron2004}
Bradley Efron and Trevor Hastie.
\newblock {Least angle regression}.
\newblock \emph{The Annals of Statistics}, pages 1--44, 2004.

\bibitem[Fledel-Alon et~al.(2011)Fledel-Alon, Leffler, Guan, Stephens, Coop,
  and Przeworski]{Fledel2011}
Adi Fledel-Alon, Ellen~Miranda Leffler, Yongtao Guan, Matthew Stephens, Graham
  Coop, and Molly Przeworski.
\newblock {Variation in human recombination rates and its genetic
  determinants.}
\newblock \emph{PloS One}, 6\penalty0 (6):\penalty0 e20321, January 2011.

\bibitem[Friedman et~al.(2010)Friedman, Hastie, and Tibshirani]{Friedman2010}
Jerome Friedman, Trevor Hastie, and Robert Tibshirani.
\newblock {A note on the group lasso and a sparse group lasso}.
\newblock \emph{arXiv preprint arXiv:1001.0736}, 2010.

\bibitem[Gabriel et~al.(2002)Gabriel, Schaffner, Nguyen, Moore, Roy,
  Blumenstiel, Higgins, DeFelice, Lochner, Faggart, et~al.]{Gabriel2002}
Stacey~B Gabriel, Stephen~F Schaffner, Huy Nguyen, Jamie~M Moore, Jessica Roy,
  Brendan Blumenstiel, John Higgins, Matthew DeFelice, Amy Lochner, Maura
  Faggart, et~al.
\newblock The structure of haplotype blocks in the human genome.
\newblock \emph{Science}, 296\penalty0 (5576):\penalty0 2225--2229, 2002.

\bibitem[Gelman et~al.(2012)Gelman, Hill, and Yajima]{Gelman2012}
Andrew Gelman, Jennifer Hill, and Masanao Yajima.
\newblock Why we (usually) don't have to worry about multiple comparisons.
\newblock \emph{Journal of Research on Educational Effectiveness}, 5\penalty0
  (2):\penalty0 189--211, 2012.

\bibitem[George and McCulloch(1993)]{George1993}
EI~George and RE~McCulloch.
\newblock {Variable selection via Gibbs sampling}.
\newblock \emph{Journal of the American Statistical Association}, 88\penalty0
  (423):\penalty0 881--889, 1993.

\bibitem[Girolami and Rogers(2006)]{Girolami2006}
Mark Girolami and Simon Rogers.
\newblock {Variational Bayesian Multinomial Probit Regression with Gaussian
  Process Priors}.
\newblock \emph{Neural Computation}, 18\penalty0 (8):\penalty0 1790--1817,
  August 2006.

\bibitem[Guan and Stephens(2011)]{Guan2011}
Yongtao Guan and Matthew Stephens.
\newblock {Bayesian variable selection regression for genome-wide association
  studies and other large-scale problems}.
\newblock \emph{The Annals of Applied Statistics}, 5\penalty0 (3):\penalty0
  1780--1815, September 2011.

\bibitem[Hans(2009)]{Hans2009}
Chris Hans.
\newblock {Bayesian lasso regression}.
\newblock \emph{Biometrika}, 96\penalty0 (September):\penalty0 835--845, 2009.

\bibitem[Hoggart et~al.(2008)Hoggart, Whittaker, {De Iorio}, and
  Balding]{Hoggart2008}
Clive~J Hoggart, John~C Whittaker, Maria {De Iorio}, and David~J Balding.
\newblock {Simultaneous analysis of all SNPs in genome-wide and re-sequencing
  association studies.}
\newblock \emph{PLoS Genetics}, 4\penalty0 (7), January 2008.

\bibitem[Ishwaran and Rao(2005)]{Ishwaran2005}
Hemant Ishwaran and J.~Sunil Rao.
\newblock {Spike and slab variable selection: Frequentist and Bayesian
  strategies}.
\newblock \emph{The Annals of Statistics}, 33\penalty0 (2):\penalty0 730--773,
  April 2005.

\bibitem[Jacob et~al.(2009)Jacob, Obozinski, and Vert]{Jacob2009}
Laurent Jacob, G~Obozinski, and JP~Vert.
\newblock {Group lasso with overlap and graph lasso}.
\newblock \emph{Proceedings of the 26th International Conference on Machine
  Learning}, 2009.

\bibitem[Jallow et~al.(2009)Jallow, Teo, and al.]{Jallow2009}
Muminatou Jallow, Yik~Ying Teo, and et~al.
\newblock {Genome-wide and fine-resolution association analysis of malaria in
  West Africa.}
\newblock \emph{Nature Genetics}, 41\penalty0 (6):\penalty0 657--65, June 2009.

\bibitem[Jeffreys(1998)]{Jeffreys1961}
Sir~Harold Jeffreys.
\newblock \emph{{The Theory of Probability}}, volume~9.
\newblock Oxford University Press, Oxford, England, 1998.

\bibitem[Jenatton et~al.(2011{\natexlab{a}})Jenatton, Audibert, and
  Bach]{Jenatton2011}
Rodolphe Jenatton, JY~Audibert, and Francis Bach.
\newblock {Structured variable selection with sparsity-inducing norms}.
\newblock \emph{The Journal of Machine Learning Research}, 2011{\natexlab{a}}.

\bibitem[Jenatton et~al.(2011{\natexlab{b}})Jenatton, Mairal, Obozinski, and
  Bach]{Jenatton2011b}
Rodolphe Jenatton, J~Mairal, Guillaume Obozinski, and Francis Bach.
\newblock {Proximal methods for hierarchical sparse coding}.
\newblock \emph{The Journal of Machine Learning Research}, 12:\penalty0
  2297--2334, 2011{\natexlab{b}}.

\bibitem[Kendziorski et~al.(2006)Kendziorski, Chen, Yuan, Lan, and
  Attie]{Kendziorski2006}
C~M Kendziorski, M~Chen, M~Yuan, H~Lan, and a~D Attie.
\newblock {Statistical methods for expression quantitative trait loci (eQTL)
  mapping.}
\newblock \emph{Biometrics}, 62\penalty0 (1):\penalty0 19--27, March 2006.

\bibitem[Kim(2009)]{Kim2009a}
Seyoung Kim.
\newblock {Tree-guided group lasso for multi-task regression with structured
  sparsity}.
\newblock \emph{Arxiv preprint arXiv:0909.1373}, 2009.

\bibitem[Kim and Xing(2009)]{Kim2009}
Seyoung Kim and Eric~P Xing.
\newblock {Statistical estimation of correlated genome associations to a
  quantitative trait network.}
\newblock \emph{PLoS Genetics}, 5\penalty0 (8), August 2009.

\bibitem[Kolter et~al.(2010)Kolter, Batra, and Ng]{kolter2010a}
J.~Z. Kolter, Siddharth Batra, and Andrew~Y. Ng.
\newblock Energy disaggregation via discriminative sparse coding.
\newblock In J.D. Lafferty, C.K.I. Williams, J.~Shawe-Taylor, R.S. Zemel, and
  A.~Culotta, editors, \emph{Advances in Neural Information Processing Systems
  23}, pages 1153--1161, 2010.

\bibitem[Kyung et~al.(2010)Kyung, Gill, Ghosh, and Casella]{Kyung2010}
Minjung Kyung, Jeff Gill, Malay Ghosh, and George Casella.
\newblock {Penalized regression, standard errors, and Bayesian lassos}.
\newblock \emph{Bayesian Analysis}, 5\penalty0 (2):\penalty0 369--412, 2010.

\bibitem[{Lango Allen} et~al.(2010){Lango Allen}, Estrada, and
  Lettre]{LangoAllen2010}
Hana {Lango Allen}, Karol Estrada, and Guillaume~\emph{et al.} Lettre.
\newblock {Hundreds of variants clustered in genomic loci and biological
  pathways affect human height.}
\newblock \emph{Nature}, 467\penalty0 (7317):\penalty0 832--8, October 2010.

\bibitem[Liu et~al.(2008)Liu, Lafferty, and Wasserman]{Liu2008}
Han Liu, John Lafferty, and Larry Wasserman.
\newblock {Nonparametric Regression and Classification with Joint Sparsity
  Constraints Joint Sparsity Constraints}.
\newblock \emph{Computer Science Department, Paper 1034}, 2008.

\bibitem[Mangravite et~al.(2013)Mangravite, Engelhardt, Medina, Smith, Brown,
  Chasman, Mecham, Howie, Shim, Naidoo, Feng, Rieder, Chen, Rotter, Ridker,
  Hopewell, Parish, Armitage, Collins, Wilke, Nickerson, Stephens, and
  Krauss]{Mangravite2013}
Lara~M. Mangravite, Barbara~E. Engelhardt, Marisa~W. Medina, Joshua~D. Smith,
  Christopher~D. Brown, Daniel~I. Chasman, Brigham~H. Mecham, Bryan Howie,
  Heejung Shim, Devesh Naidoo, QiPing Feng, Mark~J. Rieder, Yii.-Der~I. Chen,
  Jerome~I. Rotter, Paul~M. Ridker, Jemma~C. Hopewell, Sarah Parish, Jane
  Armitage, Rory Collins, Russell~A. Wilke, Deborah~A. Nickerson, Matthew
  Stephens, and Ronald~M. Krauss.
\newblock {A statin-dependent QTL for GATM expression is associated with
  statin-induced myopathy}.
\newblock \emph{Nature}, August 2013.

\bibitem[Maranville et~al.(2011)Maranville, Luca, Richards, Wen, Witonsky,
  Baxter, Stephens, and Di~Rienzo]{Maranville2011}
Joseph~C Maranville, Francesca Luca, Allison~L Richards, Xiaoquan Wen, David~B
  Witonsky, Shaneen Baxter, Matthew Stephens, and Anna Di~Rienzo.
\newblock Interactions between glucocorticoid treatment and cis-regulatory
  polymorphisms contribute to cellular response phenotypes.
\newblock \emph{PLOS Genetics}, 7\penalty0 (7), 2011.

\bibitem[Marchini et~al.(2007)Marchini, Howie, Myers, McVean, and
  Donnelly]{Marchini2007}
Jonathan Marchini, Bryan Howie, Simon Myers, Gil McVean, and Peter Donnelly.
\newblock {A new multipoint method for genome-wide association studies by
  imputation of genotypes.}
\newblock \emph{Nature Genetics}, 39\penalty0 (7):\penalty0 906--13, July 2007.

\bibitem[Meinshausen and Peter(2010)]{Meinshausen2010}
Nicolai Meinshausen and B~Peter.
\newblock {Stability selection}.
\newblock \emph{Journal of the Royal Statistical Society}, pages 1--30, 2010.

\bibitem[Mitchell and Beauchamp(1988)]{Mitchell1988}
T.~J. Mitchell and J.~J. Beauchamp.
\newblock {Bayesian Variable Selection in Linear Regression}.
\newblock \emph{Journal of the American Statistical Association}, 83\penalty0
  (404):\penalty0 1023--1032, 1988.

\bibitem[Mohamed et~al.(2011)Mohamed, Heller, and Ghahramani]{Mohamed2011}
Shakir Mohamed, Katherine Heller, and Zoubin Ghahramani.
\newblock Bayesian and $l_1$ approaches to sparse unsupervised learning.
\newblock \emph{arXiv preprint arXiv:1106.1157}, 2011.

\bibitem[Morley et~al.(2004)Morley, Molony, Weber, Devlin, Ewens, Spielman, and
  Cheung]{Morley2004}
M~Morley, C~M Molony, T~M Weber, J~L Devlin, K~G Ewens, R~S Spielman, and V~G
  Cheung.
\newblock {Genetic analysis of genome-wide variation in human gene expression}.
\newblock \emph{Nature}, 430\penalty0 (7001), 2004.

\bibitem[Murray et~al.(2010)Murray, Adams, and MacKay]{Murray2010}
Iain Murray, Ryan~P. Adams, and David J.~C. MacKay.
\newblock {Elliptical slice sampling}.
\newblock In \emph{Proceedings of the 13th International Conference on
  Artificial Intelligence and Statistics}, pages 541--548, 2010.

\bibitem[Neal(2003)]{Neal2003}
Radford~M. Neal.
\newblock Slice sampling.
\newblock \emph{The Annals of Statistics}, 31\penalty0 (3):\penalty0 705--767,
  2003.

\bibitem[O'Hara and Sillanp\"{a}\"{a}(2009)]{OHara2009}
R.~B. O'Hara and M.~J. Sillanp\"{a}\"{a}.
\newblock {A review of Bayesian variable selection methods: what, how and
  which}.
\newblock \emph{Bayesian Analysis}, 4\penalty0 (1):\penalty0 85--117, March
  2009.

\bibitem[Petretto et~al.(2010)Petretto, Bottolo, Langley, Heinig,
  McDermott-Roe, Sarwar, Pravenec, H\"{u}bner, Aitman, Cook, and
  Richardson]{Petretto2010}
Enrico Petretto, Leonardo Bottolo, Sarah~R Langley, Matthias Heinig, Chris
  McDermott-Roe, Rizwan Sarwar, Michal Pravenec, Norbert H\"{u}bner, Timothy~J
  Aitman, Stuart~A Cook, and Sylvia Richardson.
\newblock {New insights into the genetic control of gene expression using a
  Bayesian multi-tissue approach.}
\newblock \emph{PLoS Computational Biology}, 6\penalty0 (4), April 2010.

\bibitem[Polson and Scott(2010)]{Polson2010}
Nicholas~G. Polson and James~G. Scott.
\newblock {Shrink globally, act locally: sparse Bayesian regularization and
  prediction}.
\newblock \emph{Bayesian Statistics}, 2010.

\bibitem[Rasmussen et~al.(2014)Rasmussen, Hubisz, Gronau, and
  Siepel]{Rasmussen2013}
Matthew~D Rasmussen, Melissa~J Hubisz, Ilan Gronau, and Adam Siepel.
\newblock Genome-wide inference of ancestral recombination graphs.
\newblock \emph{PLoS genetics}, 10\penalty0 (5):\penalty0 e1004342, 2014.

\bibitem[Richardson et~al.(2010)Richardson, Bottolo, and
  Rosenthal]{Richardson2010}
S~Richardson, Leonardo Bottolo, and JS~Rosenthal.
\newblock {Bayesian models for sparse regression analysis of high dimensional
  data}.
\newblock \emph{Bayesian Statistics}, 2010.

\bibitem[Schwarz(1978)]{Schwarz1978}
G~Schwarz.
\newblock {Estimating the Dimension of a Model}.
\newblock \emph{The Annals of Statistics}, 6\penalty0 (6):\penalty0 461--464,
  1978.

\bibitem[Seeger(2000)]{Seeger2000}
Matthias Seeger.
\newblock {Covariance kernels from Bayesian Generative Models}.
\newblock \emph{Advances in Neural Information Processing Systems 14}, 2000.

\bibitem[Smith and Kohn(1996)]{Smith1996}
Michael Smith and Robert Kohn.
\newblock {Nonparametric regression using Bayesian variable selection}.
\newblock \emph{Journal of Econometrics}, 1996.

\bibitem[Stephens and Balding(2009)]{Stephens2009}
Matthew Stephens and David~J Balding.
\newblock {Bayesian statistical methods for genetic association studies.}
\newblock \emph{Nature Reviews Genetics}, 10\penalty0 (10):\penalty0 681--90,
  October 2009.

\bibitem[Storey and Tibshirani(2003)]{Storey2003}
John~D Storey and Robert Tibshirani.
\newblock {Statistical significance for genomewide studies.}
\newblock \emph{Proceedings of the National Academy of Sciences of the United
  States of America}, 100\penalty0 (16):\penalty0 9440--5, August 2003.

\bibitem[Storey et~al.(2005)Storey, Akey, and Kruglyak]{Storey2005}
John~D Storey, Joshua~M Akey, and Leonid Kruglyak.
\newblock {Multiple locus linkage analysis of genomewide expression in yeast.}
\newblock \emph{PLoS biology}, 3\penalty0 (8):\penalty0 e267, August 2005.

\bibitem[Stranger et~al.(2007)Stranger, Nica, Forrest, Dimas, Bird, Beazley,
  Ingle, Dunning, Flicek, Koller, Montgomery, Tavar\'{e}, Deloukas, and
  Dermitzakis]{Stranger2007}
Barbara~E Stranger, Alexandra~C Nica, Matthew~S Forrest, Antigone Dimas,
  Christine~P Bird, Claude Beazley, Catherine~E Ingle, Mark Dunning, Paul
  Flicek, Daphne Koller, Stephen Montgomery, Simon Tavar\'{e}, Panos Deloukas,
  and Emmanouil~T Dermitzakis.
\newblock {Population genomics of human gene expression.}
\newblock \emph{Nature Genetics}, 39\penalty0 (10):\penalty0 1217--1224,
  October 2007.

\bibitem[Stranger et~al.(2012)Stranger, Montgomery, Dimas, Parts, Stegle,
  Ingle, Sekowska, Smith, Evans, Gutierrez-Arcelus, Price, Raj, Nisbett, Nica,
  Beazley, Durbin, Deloukas, and Dermitzakis]{Stranger2012}
Barbara~E. Stranger, Stephen~B. Montgomery, Antigone~S. Dimas, Leopold Parts,
  Oliver Stegle, Catherine~E. Ingle, Magda Sekowska, George~Davey Smith, David
  Evans, Maria Gutierrez-Arcelus, Alkes Price, Towfique Raj, James Nisbett,
  Alexandra~C. Nica, Claude Beazley, Richard Durbin, Panos Deloukas, and
  Emmanouil~T. Dermitzakis.
\newblock {Patterns of Cis Regulatory Variation in Diverse Human Populations}.
\newblock \emph{PLOS Genetics}, 8\penalty0 (4), April 2012.

\bibitem[Tibshirani(1996)]{Tibshirani1996}
R~Tibshirani.
\newblock {Regression shrinkage and selection via the LASSO}.
\newblock \emph{Journal of the Royal Statistical Society: Series B},
  58\penalty0 (1):\penalty0 267----288, 1996.

\bibitem[Tibshirani(2014)]{Tibshirani2014}
Robert~J Tibshirani.
\newblock In praise of sparsity and convexity.
\newblock \emph{Past, Present, and Future of Statistical Science}, page 495,
  2014.

\bibitem[Tipping(2001)]{Tipping2001}
M~E Tipping.
\newblock {Sparse Bayesian learning and the relevance vector machine}.
\newblock \emph{Journal of Machine Learning Research}, 1:\penalty0 211--244,
  2001.

\bibitem[Tropp and Wright(2010)]{Tropp2010}
Joel~A Tropp and Stephen~J Wright.
\newblock Computational methods for sparse solution of linear inverse problems.
\newblock \emph{Proceedings of the IEEE}, 98\penalty0 (6):\penalty0 948--958,
  2010.

\bibitem[Valdar et~al.(2012)Valdar, Sabourin, Nobel, and Holmes]{Valdar2012}
William Valdar, Jeremy Sabourin, Andrew Nobel, and Christopher~C. Holmes.
\newblock {Reprioritizing Genetic Associations in Hit Regions Using LASSO-Based
  Resample Model Averaging}.
\newblock \emph{Genetic Epidemiology}, 12, April 2012.

\bibitem[Wilson et~al.(2010)Wilson, Iversen, Clyde, Schmidler, and
  Schildkraut]{Wilson2010}
Melanie~A. Wilson, Edwin~S. Iversen, Merlise~A. Clyde, Scott~C. Schmidler, and
  Joellen~M. Schildkraut.
\newblock {Bayesian model search and multilevel inference for SNP association
  studies}.
\newblock \emph{The Annals of Applied Statistics}, 4\penalty0 (3):\penalty0
  1342--1364, September 2010.

\bibitem[Wood et~al.(2011)Wood, Hernandez, Nalls, Yaghootkar, Gibbs, Harries,
  Chong, Moore, Weedon, Guralnik, Bandinelli, Murray, Ferrucci, Singleton,
  Melzer, and Frayling]{Wood2011}
Andrew~R Wood, Dena~G Hernandez, Michael~a Nalls, Hanieh Yaghootkar, J~Raphael
  Gibbs, Lorna~W Harries, Sean Chong, Matthew Moore, Michael~N Weedon, Jack~M
  Guralnik, Stefania Bandinelli, Anna Murray, Luigi Ferrucci, Andrew~B
  Singleton, David Melzer, and Timothy~M Frayling.
\newblock {Allelic heterogeneity and more detailed analyses of known loci
  explain additional phenotypic variation and reveal complex patterns of
  association.}
\newblock \emph{Human Molecular Genetics}, 20\penalty0 (20):\penalty0 4082--92,
  October 2011.

\bibitem[Wu et~al.(2009)Wu, Chen, Hastie, Sobel, and Lange]{Wu2009}
Tong~Tong Wu, Yi~Fang Chen, Trevor Hastie, Eric Sobel, and Kenneth Lange.
\newblock {Genome-wide association analysis by lasso penalized logistic
  regression.}
\newblock \emph{Bioinformatics (Oxford, England)}, 25\penalty0 (6):\penalty0
  714--21, March 2009.

\bibitem[Yang et~al.(2010)Yang, Benyamin, McEvoy, Gordon, Henders, Nyholt,
  Madden, Heath, Martin, Montgomery, Goddard, and Visscher]{Yang2010}
Jian Yang, Beben Benyamin, Brian~P McEvoy, Scott Gordon, Anjali~K Henders,
  Dale~R Nyholt, Pamela~A Madden, Andrew~C Heath, Nicholas~G Martin, Grant~W
  Montgomery, Michael~E Goddard, and Peter~M Visscher.
\newblock {Common SNPs explain a large proportion of the heritability for human
  height.}
\newblock \emph{Nature genetics}, 42\penalty0 (7):\penalty0 565--9, July 2010.

\bibitem[Yang et~al.(2012)Yang, Ferreira, Morris, Medland, Madden, Heath,
  Martin, Montgomery, Weedon, Loos, Frayling, McCarthy, Hirschhorn, Goddard,
  and Visscher]{Yang2012}
Jian Yang, Teresa Ferreira, Andrew~P Morris, Sarah~E Medland, Pamela A~F
  Madden, Andrew~C Heath, Nicholas~G Martin, Grant~W Montgomery, Michael~N
  Weedon, Ruth~J Loos, Timothy~M Frayling, Mark~I McCarthy, Joel~N Hirschhorn,
  Michael~E Goddard, and Peter~M Visscher.
\newblock {Conditional and joint multiple-SNP analysis of GWAS summary
  statistics identifies additional variants influencing complex traits.}
\newblock \emph{Nature Genetics}, 44\penalty0 (4):\penalty0 369--75, S1--3,
  April 2012.

\bibitem[Yuan and Lin(2005)]{Yuan2006}
M.~Yuan and Y.~Lin.
\newblock Model selection and estimation in regression with grouped variables.
\newblock \emph{Journal of the Royal Statistical Society: Series B},
  68\penalty0 (1):\penalty0 49--67, 2005.

\bibitem[Zou and Hastie(2005)]{zou2005regularization}
Hui Zou and Trevor Hastie.
\newblock Regularization and variable selection via the elastic net.
\newblock \emph{Journal of the Royal Statistical Society: Series B (Statistical
  Methodology)}, 67\penalty0 (2):\penalty0 301--320, 2005.

\end{thebibliography}

\end{document}